\documentclass[12pt,preprint]{aastex}

\newcommand{\vect}[1]{\ensuremath{\mbox{\boldmath $#1$}}}
\slugcomment{Submitted to ApJ}

\begin{document}
\def\beq   {\begin{equation}}
\def\beqa  {\begin{eqnarray}}
\def\beqan {\begin{eqnarray*}}
\def\eeq   {\end{equation}}
\def\eeqa  {\end{eqnarray}}
\def\eeqan {\end{eqnarray*}}

\def\rd { {\rm d} }
\def\rK { {\rm K} }
\def\cS { S_N}
\def\cH { {\cal{H}} }
\def\cK { {\cal{K}} }
\def\event{OGLE-1999-BUL-23}
\def\vs {\Delta u_{2,\perp}}
\def\vst {\Delta \theta_{2,\perp}}
\def\bt{\vect{\theta}}
\def\bu{\vect{u}}
\def\up{u_\perp}
\def\mtfs{{\mu_{tot}^{fs}}}
\def\btcl  { {\vect{\theta}_{cl}}  }
\def\bTcl {\delta\vect{\theta}_{cl}}
\def\bmo {{\bf 0}}
\def\det {\mathop{\rm det}\nolimits}
\def\jac {A_{\bme}}
\def\te{t_{\rm E}}
\def\thetae{\theta_{\rm E}}
\def\dos{D_{os}}
\def\dls{D_{ls}}
\def\dol{D_{ol}}
\def\cer{{\cal R}}
\def\tcc{t_{c}}
\def\mutot{\mu_{tot}}
\def\drel{D}
\def\muas{\mu{\rm as}}
\def\dtj{\Delta \theta_{jump}}
\def\grad {\vect{\nabla}}
\def\murel{\mu_{rel}}
\def\days{\rm days}
\def\msun{M_\odot}
\def\eq#1{equation~(\ref{#1})} 
\def\Eq#1{Eq.~\ref{#1}}
\def\thetas{\theta_\epsilon}
\def\hbu{\hat{\bu}}
\def\hbt{\hat{\bt}}

\newcommand{\bme}{ {\mbox{\boldmath $\eta$}} }
\newcommand{\bma}{ {\mbox{\boldmath $\alpha$}} }

\title{Gravitational Microlensing Near Caustics I: Folds}

\author{B. Scott Gaudi\altaffilmark{1,2} and A. O. Petters\altaffilmark{3,4}} 

\altaffiltext{1}{School of Natural Sciences, Institute for Advanced Study, Princeton, NJ 08540, gaudi@sns.ias.edu}
\altaffiltext{2}{Hubble fellow}
\altaffiltext{3}{Department of Mathematics, Duke University,
Science Drive, Durham, NC 27708, petters@math.duke.edu}
\altaffiltext{4}{Bass Society of Fellows, Duke University}

\begin{abstract}
We study the local behavior of gravitational lensing near fold
catastrophes. Using a generic form for the lensing map near a fold, we
determine the observable properties of the lensed images, focusing on
the case when the individual images are unresolved, i.e.,
microlensing.  Allowing for images not associated with the fold, we
derive analytic expressions for the photometric and astrometric
behavior near a generic fold caustic.  We show how this form reduces
to the more familiar linear caustic, which lenses a nearby source into
two images which have equal magnification, opposite parity, and are
equidistant from the critical curve.  In this case, the simplicity and
high degree of symmetry allows for the derivation of semi-analytic
expressions for the photometric and astrometric deviations in the
presence of finite sources with arbitrary surface brightness profiles.
We use our results to derive some basic properties of astrometric
microlensing near folds, in particular we predict for finite sources
with uniform and limb darkening profiles, the detailed shape of the
astrometric curve as the source crosses a fold.  
We find that the astrometric effects of limb darkening will be difficult
to detect with the currently planned accuracy of the {\it Space
Interferometry Mission (SIM)} for Galactic bulge sources; however, this 
also implies that astrometric measurements
of other parameters, such as the size of the source, should not be
compromised by an unknown amount of limb darkening.  We verify our
results by numerically calculating the expected astrometric shift for
the photometrically well-covered Galactic binary lensing event
OGLE-1999-BUL-23, finding excellent agreement with our analytic
expressions.  Our results can be applied to any lensing system with
fold caustics, including Galactic binary lenses and quasar
microlensing.
\end{abstract}

\keywords{astrometry---stars: binaries, fundamental parameters---gravitational lensing}

\section{Introduction}

Gravitational lensing has proven to be an exceptional tool for
studying a diverse set of astrophysical phenomena.  Its utility is
due, at least in part, to the fact it operates in a number of
qualitatively different regimes.  The term strong lensing, or
macrolensing, is usually applied when a distant source (typically
cosmological) is lensed into multiple, resolved images by an
intervening mass, such as a foreground cluster or a galaxy.  Weak
lensing is used to refer to the case when multiple images are not
created, and the gravitational field of the intervening matter serves
only to slightly distort the image of the source.  For most
applications of both strong and weak lensing, the source, lens, and
observer can be considered static.  The term microlensing is often
used to describe the case when multiple images are created, but are
not resolved.  Typically the separation of the images created by a
gravitational microlens are of order the Einstein ring radius,
\begin{equation}
\thetae=\sqrt{{{4 G M}\over \drel c^2} }.
\label{eqn:thetae}
\end{equation}
Here $M$ is the mass of the lens, $\drel$ is defined by, $\drel\equiv
\dos\dol/\dls$, and $\dos$, $\dol$, and $\dls$ are the distances
between the observer-source, observer-lens, and lens-source,
respectively.  In cosmological contexts, angular diameter distances
should be used.  When $\thetae$ is less than the resolution,
individual images in general cannot be distinguished.  Due to the
small scale of $\thetae$, it is typically not a good approximation to
assume that the source, lens, and observer are static.  Therefore the
lensing properties can be expected to change on timescales of order
the Einstein ring crossing time,
\begin{equation}
\te = {\thetae \dol\over v_\perp},
\label{eqn:te}
\end{equation}
where $v_\perp$ is the transverse speed of the lens relative to the observer-source
line-of-sight.  The
standard observables in gravitational microlensing are therefore
the time rate of change of the total magnification and center-of-light
(centroid) of all the microimages.  There are two different regimes
where microlensing has been discussed: quasar microlensing \citep{wambs2001}
and microlensing in the Local Group \citep{pac1996}.

In the Local Group, gravitational microlensing occurs whenever a massive, compact
object passes close to our line of sight to a more distant star.  Microlensing was
originally suggested as a method to detect baryonic dark matter in the
halo of our galaxy \citep{pac1986}, but has been developed and applied
as an important tool in studying a number of astrophysical topics,
including the stellar mass function \citep{gould1996}, extrasolar
planets \citep{mandp1991}, stellar atmospheres \citep{gould2001}, and
stellar multiplicity \citep{alcock2000,udalski2000}.  The only
microlensing effect currently observable is the magnification of the
background source as function of time.  This is because, for typical 
distances in the Local Group, the angular
Einstein ring radius is $\thetae\simeq 1~{\rm mas}({M / \msun})^{1/2},$ 
and therefore too small to be resolved with current instruments.  
The timescale for a microlensing event is $\te  \sim 40~\days$. 
In general, it is much easier to determine the center-of-light of an image 
than it is to resolve it.  Thus 
future interferometers, such as
the {\it Space Interferometry Mission (SIM)}, although still not able
to resolve separations of ${\cal O}(\rm mas)$, should be able to
measure the centroid of all the images to much better than this, perhaps
even down to $10~\muas$ in the case of {\it SIM}.  
Such accuracy is sufficient to
easily detect the motion of the centroid of the images created in a
microlensing event, which is also of order $\thetae$.  This regime is
typically referred to as astrometric microlensing, as opposed to
photometric microlensing when only the total magnification is
observable.  

Astrometric microlensing has a number of important 
applications.  By combining ground-based photometry
of microlensing events with photometry and astrometry
from an astrometric satellite on an Earth-like orbit,
the masses of microlenses can routinely be measured 
\citep{pac1998, bsvb1998, gs1999}, allowing for 
the determination of the compact object mass
function in the bulge, including stellar remnants \citep{gould2000}. 
Astrometric information alone allows for the
precise (few \%) measurement of the masses of nearby stars
\citep{pac1995}.  Finally, for a subset of 
events, it will be possible to obtain precision measurements
of angular diameters of stars in the Galactic bulge using
astrometric information \citep{pac1998,ggh2002}.

Binary microlenses have proven to be enormously useful in photometric
microlensing studies.  This is primarily because binary lenses exhibit
caustics: closed curves on which the mapping from the lens plane to light source 
plane becomes critical,
and the point-source magnification becomes
formally infinite.  Regions near caustics exhibit large,
rapidly changing (with respect to source position) magnification, and
are therefore useful not only for providing a large source flux, but
also high angular resolution.  However, binary lenses have also proven
to be difficult to study both theoretically and observationally.  This
is partially because the lens equation, which describes the mapping
from the lens plane to the light source plane, is equivalent to a fifth-order
complex polynomial in the source position \citep{witt1990}, and
therefore is not analytically solvable in general.  Furthermore, care
must be taken when considering finite source effects near caustics due
to the divergent magnification.  However, considerable progress can be
made when one realizes that the smooth arcs (away from cusps)
of caustics
that arise in nearly equal-mass binary lenses are well approximated as
simple linear fold catastrophes, which have generic, simple, and most
importantly, {\it analytic} behavior.  Thus the caustics of binary
lenses can be analyzed analytically or semi-analytically without
reference to the global (and non-analytic) topology of the general
binary lens.  In particular, a simple equation for the magnification
of a source near a fold exists \citep{sef92}, which has been used in a
number of important applications including binary-lens fitting
\citep{albrow1999}, stellar atmospheres \citep{gandg1999}, and
caustic-crossing predictions \citep{jandm2001}.

In contrast to the astrometric behavior of single lenses, which is
analytic and has been quite well studied \citep{walker1995,jhp1999},
there have been only a few preliminary studies of the
astrometric properties of binary gravitational lenses \citep{hcc1999,
ch1999,gh2000}.  It is known that astrometric
binary-lens curves exhibit complex behavior, including instantaneous
${\cal O}(\thetae)$ jumps in the image centroid trajectory that occur
when a point source crosses a binary-lens caustic and two
highly-magnified images appear in a position unrelated to the position
of the centroid of the other three binary-lens images.  
The generic behavior of these centroid jumps, or how they are altered by
finite source effects, is not understood.  As is the case for
photometric microlensing, the astrometric behavior of binary lenses
will likely prove quite useful for several applications.  The
usefulness of binary lenses is primarily related to the complex image
centroid trajectories and the large centroid jumps.  Although, in
general, these properties do not allow one to measure any additional
parameters over the single lens case; they do allow one to measure
these parameters much more easily.  In particular, \citet{gg2002}
have shown that lens mass measurements can be made to a given
accuracy with 1-3 orders of magnitude fewer photons with caustic-crossing binary-lens
events than with single lens
events, thus greatly reducing the resources required to achieve one of
the primary proposed science goals of astrometric microlensing.
Caustic crossing binary-lens events are also enormously useful for
measuring the angular radii of microlensing source stars in the bulge, for two
reasons.  First, the expected ratio of binary-to-single lens
events for which the source star is resolved is a factor of $\ga 4$
for giant sources and $\ga 10$ for main
sequence sources.  Furthermore, the large and complex centroid shifts
expected for caustic-crossing binary-lens events makes the requisite astrometric
measurements easier.  \citet{ggh2002} have shown that,
by combining accurate ground-based photometry with a handful of precise
astrometric measurements, caustic-crossing
binary-lens events should yield $\sim 5\%$ stellar radius measurements with
reasonable expenditure of resources.\footnote{Although it is possible to measure
stellar angular sizes using astrometric information alone, this generally requires
very densely-sampled measurements, since the source is resolved
only for a short time.}
Thus, given the importance of caustic crossing binary-lens events,
an analytic study of the generic behavior
of astrometric microlensing near folds would prove quite useful.

In quasar microlensing, the separate macroimages of a quasar that is
multiply-imaged by a intervening galaxy or cluster also feel the
combined, non-linear effect of individual point masses (i.e. stars) in
the macrolensing object that are near the macroimage position.  The
individual macroimages are in fact composed of many, unresolved
microimages with separations of order the Einstein radius of a $M\sim
\msun$ object at cosmological distances, $\thetae \simeq 1
\muas(M/\msun)^{1/2}$.  The typical timescale for the source to cross
an angle of $\thetae$ is $\te \simeq 15~{\rm years}$; however
microlensing light curves should show structure on much smaller
timescales due to the combined effects of many individual microlenses.
Since it was first discussed by \citet{cr1979}, cosmological
microlensing has been studied theoretically by numerous authors (see
\citealt{wambs2001} and references therein), and detected in at least
two systems (Q2237+0305, \citealt{irwin1989,corrigan1991,wozniak2000},
B1600+434, \citealt{koop2000}).  Observations have been used to place
constraints on, e.g.,\ the size of the emitting region of quasars
\citep{wambs1990, wyithe2000b} and the mass function of microlenses
\citep{schmidt1998,wyithe2000a, koop2000}.

Quasar microlensing differs markedly from microlensing in the Local
Group in that the surface mass density in units of the critical
density for lensing (the ``optical depth'') is of order unity, rather
than ${\cal O}(10^{-6})$ for the Local Group.  In the high optical
depth regime the lensing effects of the individual microlenses add
nonlinearly, resulting in a complex caustic network.  Due to this
nonlinear behavior and the large number of lenses typically involved,
calculation of the observable properties of such a lensing system is
difficult and time consuming.  Although in the high optical depth
regime the caustics often exhibit considerably more complicated
global behavior than the caustics of binary lenses in the Local Group,
it is still the case that the  
smooth arcs (away from cusps) of the caustics
are locally well-approximated by generic fold catastrophes.
This fact, combined with a simple formula for the magnification near folds,
has been exploited by numerous authors to quickly and efficiently calculate 
various observable properties of quasar microlensing 
\citep{wambs1991,lw1998,ww1999,fw1999}.  

The observable effects of quasar microlensing have been limited to the
relative magnifications of the various macroimages as a function of
time.  As with Local Group microlensing, astrometric effects should
also be present.  The centroid of the individual
macroimages should vary as a function of time, particularly when new images 
are created or destroyed when the source crosses a caustic.  This effect has been
studied by \citet{ws1995} and \citet{lw1998}.  In particular,
\citet{lw1998} predict that magnitude of the centroid shift for the
Q2237+0305 system 
(apparent magnitude $R\la 18.5$)	
can be as large as $\sim 50\muas$, and thus potentially 
observable with {\it SIM}.  They also note that the magnitude of the
centroid shift is often correlated with the magnitude of the change in
total magnification.  The analytic results presented in this paper
may prove useful for this application.

Here we study the generic, local behavior of microlensing near fold
catastrophes.  In \S 2 we present an analytic study of the photometric
and astrometric behavior near folds.  We begin with the equations that
describe the mapping near a fold caustic in \S 2.1, and use these to
derive the behavior of a point source near a fold.  We extend this
analysis to finite sources in \S 2.2, and limb-darkened sources in \S
2.3.  In \S 2.4 we show how and when our generic parabolic fold form
reduces to the more familiar linear caustic.  In \S 2.5 we use our
analytic results to derive some generic results about the astrometric
behavior near folds.  We verify the applicability of our results in \S
3 by numerically calculating the photometric and astrometric behavior
of one well-observed binary-lens event.  We find excellent agreement
with our analytic formulae.  Finally, we summarize and
conclude in \S 4.

Our goal is to provide a thorough, comprehensive study of
gravitational microlensing near fold caustics.  Although our
study is interesting in its own right, the primary utility of the
results presented here is their potential application to the topics
mentioned in the previous paragraphs.  A prescription for how
specifically our results can be applied to these topics is beyond the scope of
this paper, but we will make general comments along these lines over
the course of the paper, and more specific comments in \S 2.5.  We are currently
preparing a complementary, similarly detailed study of microlensing
near cusps.  Combined with this study, we will have a reasonably
thorough and complete understanding of the local behavior of
microlensing observables near all stable gravitational lensing singularities.
We note that some of the results derived here, particularly the
results on the photometric behavior near folds, have been presented
elsewhere (see, e.g. \citealt{sef92}, \citealt{plw01}, and
\citealt{fw1999}).  We include those results here for the sake of
completeness.

\section{Analytic Considerations \label{sec:analytic}}

\subsection{Lensing Near Fold Caustics}

\subsubsection{Global Lensing Map}

For a general gravitational lens, the lensed images $\bt$ of
a source $\bu$, are given by the solutions of the lens equation,
which is the mapping $\bu \rightarrow \bt$:
\begin{equation}
\bu = \bme (\bt) \equiv \bt - \bma (\bt),
\label{eqn:lenseq}
\end{equation}
where $\bma= \grad \psi$, and $\psi$ is the projected Newtonian potential of
the lens,
\begin{equation}
\psi(\bt)= {1 \over \pi } \int_{\rm R^2}d{\bt'} \kappa(\bt')\ln |\bt-\bt'|.
\label{eqn:psidef}
\end{equation} 
Where $\kappa=\Sigma(\bt)/\Sigma_{cr}$, $\Sigma(\bt)$ is the surface density of
the lens, and 
\begin{equation}
\Sigma_{cr}\equiv {c^2 \over 4 \pi G} {\dos \over \dol \dls}
\label{eqn:sigcrit}
\end{equation}
 is the critical surface density for lensing.  Note that $\grad^2 \psi=2\kappa$. 
We are assuming that $\bt = \vect{r}/({\theta_{\rm E} \dol})$ and 
$\bu = \vect{s}/({\theta_{\rm E} \dos})$, 
where $\vect{r}$ and $\vect{s}$ are the proper vector
positions in the lens and light source planes, respectively.  
The mapping $\bme$ can produce multiple images of the source $\bu$.
The magnification of lensed image $\bt_i$ is
\begin{equation}
\mu (\bt_i) = \frac{1}{|\det [A_\bme(\bt_i)]|},
\label{eqn:magoneimage}
\end{equation}
where $\jac$ is the Jacobian matrix of the lensing map $\bme$.
The critical curve is the set of positions $\bt_c$ such that the
determinant of the Jacobian matrix vanishes, i.e.\ where 
\begin{equation}
J\equiv \det[A_\bme(\bt_c)]=0,
\label{eqn:jdef}
\end{equation}
and the caustics are $\bu_c=\bme
(\bt_c)$.  In the case of microlensing, the individual images
are by definition unresolved, and
thus it is useful to define the total magnification
\begin{equation}
\mutot = \sum_{i} \mu(\bt_i),
\label{eqn:mutot}
\end{equation}
where the sum is over all images.
The center-of-light, or centroid $\btcl (\bu)$ of the images is simply
the magnification weighted sum of the image positions,
\begin{equation}
\btcl (\bu) = {{\sum_i \mu(\bt_i)\bt_i} \over {\mutot}}.
\label{eqn:cl}
\end{equation}
For simplicity, we will focus primarily on the quantity $\btcl$. 
However, it is important to note that, in general, centroid measurements will be made 
with respect to the unlensed source position, thus the observable is
\begin{equation}
\bTcl (\bu) = \btcl (\bu) - \bu.
\label{eqn:dcl}
\end{equation}
Also, while the angular variables
we will be working with will be in units of $\thetae$, the astrometric observables are
in physical units, such as arcseconds.  To convert to observable quantities, 
all angular quantities must be multiplied by $\thetae$.  For example,
the observable centroid shift is given by $\delta \vect{\varphi}_{cl}\equiv\thetae \bTcl$.

\subsubsection{Lensing Map Near Folds}  

We now derive the generic behavior of the photometric and
astrometric properties of gravitational lensing near folds.  We will present our derivations
in some detail, in order to document the approximations and simplifying assumptions
that are implicit in the final analytic expressions.   In Figure 1, we provide an illustrative
example of the basic properties of lensing near a fold.  We will refer
to this figure repeatedly during the course of the derivations. 

Suppose that the lensing map $\bme$ sends the origin to itself (which
can always be accomplished by appropriate translations) and
a fold caustic curve
passes through the origin.  By Taylor expanding the gravitational potential
$\psi$ about the origin, one can find 
an orthogonal change of coordinates that is the same in the lens
and light source planes such that the lensing
map $\bme$ can be approximated by the following mapping
in a neighborhood of the origin
(Petters, Levine, \& Wambsganss 2001, pp. 341-353; 
Schneider, Ehlers, \& Falco
1992, p. 187):
\beq
\label{eq-le-fold}
u_1 = a  \theta_1 + \frac{b}{2}   \theta^2_2  +   c  \theta_1 \theta_2,
\qquad 
u_2 =  \frac{c}{2}  \theta^2_1 + b  \theta_1  \theta_2 + \frac{d}{2} \theta^2_2,
\eeq
where $(\theta_1, \theta_2)$ and $(u_1,u_2)$ denote the respective coordinates
in the lens and light source planes, and
\beq
a = 1 - \psi_{11} (\bmo) \neq 0,
\quad b = -\psi_{122} (\bmo),
\quad c = - \psi_{112} (\bmo),
\quad d = - \psi_{222} (\bmo) \neq 0.
\eeq
Here the subscripts refer to the partial derivatives  of $\psi$  with respect to the 
original global Cartesian coordinates of the
lensing map.  For the example in Figure 1, we have adopted $a=5$, $b=1$, and 
$c=-d=-0.5$. 

The Jacobian matrix of \eq{eq-le-fold} is
\beq
 A =
\left [
\begin{array}{cc}
a + c \theta_2  & c \theta_1 + b \theta_2\\ \\
c \theta_1 + b \theta_2 &  b \theta_1 + d \theta_2
\end{array}
\right ].
\eeq
The critical curve
is given by
\beq 
J \equiv \det A = (a + c \theta_2)(b \theta_1 + d \theta_2) - 
(c \theta_1 + b \theta_2)^2 = 0.
\eeq
The tangent line to the critical curve at the origin is given by
\beq
\label{eq-gradJ}
0 = \bt \cdot \grad J (\bmo)
= ab \theta_1 + ad  \theta_2,
\eeq
that is,
\beq
\label{eq-tanline-cc}
\theta_2 =  - \frac{b}{d} \theta_1
\eeq
since $a \neq 0$ and $d \neq 0$. Substituting $\theta_2 =  - b\theta_1/d$ into
\eq{eq-le-fold} yields
\beq
u_1 = a  \theta_1 + \frac{b}{d^2} \left(\frac{1}{2}b^2 - cd\right)  \theta^2_1
\simeq a  \theta_1, \qquad 
u_2 = \frac{1}{2d}(cd - b^2)\ \theta^2_1.
\eeq
Note that in the expression for $u_1$, the
term $\theta_1$ dominates $\theta_1^2$ near the origin.
Inserting $\theta_1 = u_1/a$ into $u_2$ above, we see that
the tangent line at the origin of the critical curve is
mapped into a parabola \citep{sef92,fw1999}: 
\beq
C (u_1,u_2) \equiv  2a^2d(u_2 -e u_1^2)= 0.
\label{eq-C}
\eeq
where we have introduced the combination of local derivatives of $\psi$:
\beq
e \equiv \frac{cd -b^2}{2a^2d}.
\label{eq-curvature-e}
\eeq
We show in \S\ref{sec:simple} that $|e|$ is one-half the curvature of the caustic at the
origin.  Thus when $|e| \ll 1$, the caustic can be approximated as $2a^2d u_2=0$, i.e.\ the
$u_1$-axis.  For the example shown in Figure 1, $|e|=0.05$.

Since the tangent line (\Eq{eq-tanline-cc}) approximates
the critical curve near the origin,  the parabola 
(\Eq{eq-C})  
approximates the caustic near the origin.  See Figure 1(a,b).
Now multiplying $u_2$ in \eq{eq-le-fold} by $2 a^2$
and substituting  $\theta_1 = u_1/a$ 
yields
\beq
\label{eq-C-2}
C (u_1,u_2) = 
(b \ u_1 \ + \ ad \ \theta_2)^2.
\eeq
Hence, if a light  source is located at a position $(u_1, u_2)$
where $C (u_1, u_2) < 0$, then there is no lensed image locally,
while a source with $C (u_1, u_2) \ge 0$  has at least one image.

For an $n$-point mass lens, we have $ c =- d$.  Consequently,
if $d >0$, the parabolic caustic lies in the lower-half
plane locally.  In addition,
the region above the parabola is such that sources lying there
have double images locally (Fig. 1a).  If $d <0$, then parabola is in
the upper-half plane with the region below the parabola yielding
double images locally.  In other words, the caustic is locally
convex (see Petters et al. 2001, Sec 9.3 for a detailed treatment).

\subsubsection{Image Positions of Sources Near Folds}

Let us determine the images for $C (u_1, u_2) \ge 0$.
Equation (\ref{eq-C-2}) is equivalent to a quadratic
equation in $\theta_2$, 
\beq
0  = [a^2 d^2] \ \theta_2^2 \ +  \ [2abd u_1] \ \theta_2
      +  [b^2  u_1^2 - C(u_1,u_2)].
\eeq
The solutions are\footnote{Note that we do not need to include ${\rm sign} (ad)$
in front of the square root since $\pm {\rm sign} (ad) = \pm 1$.}
\beq
\label{eq-image-2}
\theta_2 = \frac{-b u_1 \ \pm \sqrt{C(u_1,u_2)}}{ad}.
\eeq
The expression for $u_1$ in \eq{eq-le-fold} 
yields
\beq
\label{eq-th1}
\theta_1 =  \frac{1}{a  +  c \theta_2}
             \left( u_1  -  \frac{b}{2}  \theta^2_2 \right).
\eeq
Ignoring terms of order 3 or higher, we obtain
\beq
\theta_1 
 = \frac{u_1}{a} - \frac{c}{a^2} u_1  \theta_2 - \frac{b}{2a}  \theta_2^2,
\label{eq-th1-2}
\eeq
where the approximation $(a \ + \ c \theta_2)^{-1}
\simeq a^{-1} (1 - c \theta_2/a)$ was employed. 
Substituting \eq{eq-image-2} into 
(\ref{eq-th1-2}) 
and keeping only terms that are linear in $u_1$ and $u_2$, it follows that
\beq
\label{eq-image-1}
\theta_1 = \frac{d u_1-b u_2}{ad}.
\eeq
Hence, a source with $C (u_1,u_2) > 0$ has two (opposite parity) images
given by
\beq
\label{eq-image-positions}
\bt_\pm \equiv (\theta_{\pm,1}, \theta_{\pm,2})
= \frac{1}{ad} \left( d u_1  \  - \  b u_2,  \ 
                -b u_1 \ \pm \sqrt{C (u_1,u_2)} \right).
\eeq
Figure 1(a,b) illustrates the mapping from source to images near a fold. 

Consider a point $\bu_c$ on the caustic, i.e., $C (\bu_c) =0$.  
Locally there is one image of $\bu_c$, located on the critical curve
\footnote{That the image position in \eq{eq-image-cc} corresponds to a point
on the critical curve can be seen by noting that 
on the caustic $u_{c,2}=e u_{c,1}^2$, 
and close to the origin, $u_1$ dominates $u_1^2$, and thus the image
position is $\bt_c =(d u_{c,1},-b u_{c,1})/(ad)$, which is indeed a 
point on the critical line 
(see \Eq{eq-tanline-cc}).}
at the position
\beq
\bt_c 
= \frac{1}{ad} (d u_{c,1}  -   b u_{c,2}, -b u_{c,1}).
\label{eq-image-cc}
\eeq
Now consider a source at $\bu = (u_1, u_2)$
inside the caustic (i.e., $C(\bu) >0$)  and
let $\bu_c^* = (u^*_{c,1},u^*_{c,2})$ 
be the point on the caustic with the same horizontal position as $\bu$,
i.e., $u^*_{c,1} = u_1$ and $u^*_{c,2} \neq u_2$.
The vertical separation of the source from the caustic
is 
then the difference 
between the $u_2$-coordinates of 
$\bu$ and $\bu^*_c$:
\beq
\label{eq-vsepartion}
\vs \equiv u_2  -  e (u^*_{c,1})^2 
    = \frac{C(u_1,u_2)}{2 a^2 d}.
\eeq
Note that $\vs > 0$ if and only if $d >0$ (since the source is located where
$C(u_1,u_2) >0$).
Using equations (\ref{eq-image-positions}) 
and (\ref{eq-image-cc}), we find that the distances of the images 
$\bt_\pm (u_1,u_2)$ from the point $\bt_c$ on the critical line are the same,
\beq
\label{eq-image-distance}
\vst=\left| \bt_\pm (u_1,u_2) - \bt_c \right| = 
\sqrt{\frac{2\vs}{d}} \left[1+\frac{b^2}{2a^2d}\vs\right]^{1/2}.
\eeq
Near the origin, the first term in brackets dominates over the second, and thus,
\beq
\label{eq-absimage-distance}
\vst= \sqrt{2\vs\over d}.
\eeq

\subsubsection{Magnification of Sources Near Folds}

Using the earlier tangent line approximation to
 the critical curve at the origin,
we Taylor expand the Jacobian determinant $J$ about the origin to
first order:
\beq
J(\theta_1, \theta_2) = J_{\theta_1} (\bmo) \theta_1  \ +  \ 
J_{\theta_2} (\bmo) \theta_2
= ab \ \theta_1 \ + \ ad \ \theta_2,
\eeq
which is equal to the right-hand-side of (\ref{eq-gradJ}).  
Inserting the expressions for $\theta_1$ and $\theta_2$ from 
\eq{eq-image-positions}, we obtain
\beq
J(\theta_1, \theta_2) = -\frac{b^2}{d}u_2 \pm \sqrt{C(u_1,u_2)}.
\eeq
Keeping only the lowest order terms in $u_1$ and $u_2$ gives,
\beq
|J(\bt_\pm)| = \sqrt{{C(u_1,u_2)}}.
\eeq
In terms of the vertical separation between the source and
the caustic (\Eq{eq-vsepartion}), the magnification of the images are given locally
by
\beq
\label{eq-mag}
\mu_\pm \equiv \mu (\bt_\pm) = \frac{1}{|J(\bt_\pm)|}
= \frac{1}{2} \sqrt{\frac{u_{f}}{\vs}},
\eeq
where
$u_{f} =   2/(a^2 d)$.  Thus the two images have the same magnification
if the source is sufficiently close to the caustic.  The total magnification $\mu_f\equiv 2\mu_\pm$ of the two 
images is simply 
\beq
\mu_f = \sqrt{\frac{u_{f}}{\vs}},
\label{eq-muf}
\eeq
which agrees with the expressions derived by both Schneider, Ehlers \& Falco (1992, p190) 
and \citet{fw1999}. 
Equation (\ref{eq-muf}) is the well-known result that the
magnification varies inversely as the square-root of the distance of
the source from the caustic.  It is not often appreciated however,
that this property holds for the general parabolic fold form as well,
provided that the {\it vertical distance} is
used, rather than, for example, the minimum distance between the
source and the fold caustic.  Rigorously, the vertical distance is the 
distance between the source and the caustic in the direction perpendicular to the
tangent line of the caustic at the origin, where the origin is defined
as the point around which the potential $\psi$ is Taylor expanded. 
In the limit of a straight fold ($|e| \ll 1$), the vertical distance and the minimum distance
are equivalent.

\subsubsection{Image Centroid of Sources Near Folds}

Write the source position $\bu$ in terms of the fold local
coordinates, i.e.,  $\bu = (u_1, u_2)$.
There are no local images for source a position with $ C (\bu) < 0$.  From
equations (\ref{eqn:cl}) and (\ref{eq-image-positions}), and since $\mu_+=\mu_-$, for
sources inside the caustic the centroid is given locally as 
\beq
\label{eq-centroidf}
\bt_{f} (\bu) 
=  \frac{1}{2}\left( \bt_+ + \bt_-\right)
= \frac{1}{ad} (d u_1  \ - \ b u_2, \ - b u_1),
\qquad \quad {\rm for} \ 
\ C (\bu)\ge 0.
\eeq
We express the rectilinear motion of the source as follows:
\beq
\label{eq-trajectory}
\bu (t) = \bu_c +  (t - t_c) \dot{\bu},
\eeq
where $\bu_c$ is the position at which the source intersects
the caustic at time $t=t_c$ and $\dot{\bu}$ is the constant angular velocity vector
of the source,
\beq
\dot{\bu}= \left(
\frac{\cos \phi }{\te}, \frac{\sin \phi}{\te}\right).
\eeq
Recall that $\te = {\thetae \dol/v_\perp}$,
where $v_\perp$ is the transverse speed of the lens relative to the observer-source
line-of-sight.  
Here $\phi$ is the angle of the source's trajectory with respect to the $u_1$-axis.
Note that the $u_1$-axis does not necessarily coincide with the caustic, and thus
$\phi$ is not necessarily the angle of the trajectory with respect to the tangent
to the caustic at $\bu_c$. 

We shall assume that $t > t_c$ corresponds to the source's trajectory
lying in the double-image region.
In that region, the centroid follows a straight line locally:
\beq
\label{eq-centroidf2}
\bt_f (t)   
=  \bt_{f,c} + (t - t_c) \dot{\bt}_{(f,c)}, \qquad \quad {\rm for} \ t \ge t_c
\eeq
where $\bt_{f,c} \equiv \bt_f (\bu_c)$
and $\dot{\bt}_{(f,c)} \equiv {\rm d} \bt_f/{\rm d} t|_{t=t_c} =  \bt_f (\dot{\bu})$
(both constant vectors).
Note that since the slope of the centroid line
$\bt_f$ is,
$\tan \phi_c = (-b \cos \phi)/(d \cos \phi - b \sin \phi)$,
where $\phi_c$ is the angle between the centroid line and
the $\theta_1$-axis, it follows that $\dot{\bt}_{(f,c)}$
can be expressed as
\beq
\dot{\bt}_{(f,c)} =  \frac{-b \cos \phi}{a d \ \te} (\cot \phi_c , 1).
\eeq

\subsubsection{Global Magnification and Centroid}

Fold caustics do not, of course, exist in isolation.  They are tied to
the global properties of the lens in consideration.  For practical
purposes, we therefore consider images created by the lens that are
not associated with the fold under consideration.  We define $\mu_0$
to be the total magnification of all the images not associated with
the fold, and we define $\bt_0$ as the centroid of all these images.
We will assume that there is locally only one fold
caustic, and that all the other image magnifications and positions are
only slowly varying functions of the source position.  The total
magnification is then
\begin{equation}
\mutot=\mu_{f}+\mu_0  =  \left( \frac{u_{f}}{\vs}\right)^{1/2} \ \Theta (\vs) \ + \ \mu_0.
\label{eqn:mutotexp}
\end{equation}
See Figure 1(c). 
Here $\Theta$ is the Heaviside unit step function
(i.e., $\Theta (x) = 1$ for $x \ge 0$ and
 $\Theta (x) = 0 $ if $x  < 0$);  it accounts for the fact
that sources below the caustic (with $\vs < 0$) have no images locally (i.e., near the critical curve).
Since
$\mu_+=\mu_-= \mu_{f}/2$, we have,
\beq
\label{eqn:genbxcl}
\btcl = {1 \over \mutot}[\mu_{f} \bt_{f} + \mu_0 \bt_0]
\eeq
where 
\beq
\bt_{f} (\bu)  
= \frac{1}{ad} (d u_1- b u_2,- b u_1) \Theta (\vs).
\eeq

We can calculate the dependence of the observables $\mutot$ and $\btcl$ on
time by assuming a rectilinear source trajectory and
replacing $\bu$ in equations (\ref{eqn:mutotexp}) and (\ref{eqn:genbxcl}) by
\eq{eq-trajectory}.
We first Taylor expand $\bt_{0}$ and $\mu_0$ about the time of the caustic crossing,
$t=\tcc$, keeping terms of first order in $t - t_c$:
\beqa
\bt_0  & = &
\bt_{0,c} + (t-\tcc) \dot{\bt}_{(0,c)},
\label{eqn:btfvt} \\
\mu_0 & = & \mu_{0,c} + (t-\tcc) \dot{\mu}_{(0,c)},
\eeqa
where $\dot{\bt}_{(0,c)} \equiv {\rm d}\bt_0 /{\rm d}t|_{t=t_c}$ 
and $\dot{\mu}_{(0,c)}
     \equiv {\rm d}\mu_0 /{\rm d}t|_{t=t_c}$. 
From equations 
(\ref{eq-vsepartion}) and  (\ref{eq-trajectory}), 
$\vs$ is given by 
\beq
\vs= \frac{t-\tcc}{\te} \sin\phi \left[ 1- e \cot\phi \left(
       2 u_{c,1} + \frac{t-\tcc}{\te} \cos\phi \right) \right].
\eeq 
The magnification as a function of time is then,
\begin{eqnarray}
\mu_{tot} (t) 
& = & \left(\frac{t_{f}}{t-\tcc}\right)^{1/2}
\left[ 1-  e \cot\phi \left( 2 u_{c,1} +  \frac{t-\tcc}{t_f} u_f \cot\phi \right)\right]^{-1/2} \Theta (t-\tcc)  
\nonumber \\
& & \hspace{3.5in}
+ \   \mu_{0,c} \  + \ 
         \dot{\mu}_{(0,c)} (t-\tcc),
\nonumber \\
\label{eqn:muvtime}
\end{eqnarray}
where 
$t_{f}\equiv (u_{f} \te)/\sin \phi$
is the effective rise time of the caustic crossing, and 
$u_{f} = 2/(a^2 d)$ defines the characteristic rise length as before.
Notice that when the curvature of the caustic is small, i.e.,\ when $|e| \ll 1$, the
magnification associated with the fold reduces to the more familiar form
$\mu_f = [(t-t_c)/t_f]^{-1/2}\Theta(t-t_c)$.
The centroid of all images is
\begin{equation}
\btcl(t) = \frac{1}{\mutot}
\left\{ 
[\mu_{f} \bt_{f,c}\Theta (t-\tcc) + \mu_0 \bt_{0,c}]+
[\mu_{f}\dot{\bt}_{(f,c)}\Theta (t-\tcc) + \mu_0 \dot{\bt}_{0,c}] 
(t-\tcc) \right\},
\label{eqn:bxclvtime}
\end{equation}
where $\bt_{f,c}$ and $\dot{\bt}_{(f,c)}$ are defined below \eq{eq-centroidf2}.  Figure 1(e,f) illustrates
the behavior of the two components of $\btcl(t)$ as a function of time, whereas Figure 1(g) shows
$\theta_{cl,1}$ versus $\theta_{cl,2}$.

\subsection{Finite Sources\label{sec:fsources}}

\subsubsection{Finite Source Magnification}

The results of the previous section assumed a pointlike source.
This results in an astrometric curve that exhibits an
instantaneous jump from $\bt_{0,c}$ to $\bt_{f,c}$ at
$t=\tcc$.  All real sources will have a finite extent which
will smooth out the discontinuous jump.  For a finite source, the
magnification is the surface brightness weighted magnification
integrated over the area of the source,
\begin{equation}
\mu^{fs}={{\int_D {\rm d} \bu S(\bu) \mu(\bu)}\over{\int_D {\rm d} \bu S(\bu)}},
\label{eqn:fsgen}
\end{equation}
where $D$ is the disc-shaped region of the source and
$S(\bu)$ is the surface brightness of the source.  
Let $\bar{S}$ be the average surface brightness of the source,
$\bar{S} \equiv (\pi \rho^2_*)^{-1}\ \int_D {\rm d}\bu S(\bu)$.
Here $\rho_*\equiv\theta_*/\thetae$, is the angular source radius
$\theta_*$ in units of $\thetae$.  The denominator in \eq{eqn:fsgen} is then 
simply $\pi\rho_*^2 \bar{S}$. Define 
$\cS\equiv S(\bu)/{\bar{S}}$ to be the normalized surface brightness. 
Also define a new set of source plane coordinates such that
\beq
\bu' = \frac{\bu - \bu_{cn}}{\rho_*},
\eeq
where $\bu_{cn}$ is the position of the center of the source.
Then for a point $\bu$ inside the disc source $D$, we have $|\bu'| \le 1$.  
Equation (\ref{eqn:fsgen}) becomes  
\beq
\mu^{fs}= \frac{1}{\pi}\int_D {\rm d}\bu' \cS(\bu') \mu(\bu').
\label{eq-mufsscale}
\eeq

All of the preceding results apply to generic parabolic fold
catastrophes.  However, in order to continue making significant
progress analytically, we must make the following simplifying
assumption: We furthermore assume that $|e| \ll 1$, and thus the
caustic coincides with the $u_1$-axis and $\vs = u_2$.  This
considerably simplifies the form for the fold magnification $\mu_f$.

Let $z$ be such that $z \rho_*$ is the vertical separation of the
center $\bu_{cn}$ of the source from the caustic, i.e., $u_{{cn},2} =
z \rho_*$.  The source $D$ is in the upper-half plane (i.e., $u_2 \ge
0$) if and only if $z\ge 1$.  If $z=1$, then $D$ just touches the
caustic, while for $0 < z < 1$ a portion of $D$ is below the caustic
with the center of $D$ on the caustic for $z=0$.  The center of $D$
lies below the caustic for $z<0$, with a portion of $D$ still above
the caustic for $ -1 < z < 0$ and $D$ completely below the caustic for
$z<-1$.  Since the fold magnification $\mu_f$ is nonzero only for
points $\bu$ in $D$ that lie in the upper-half plane, we get
\beq
\label{eq-mufp}
\mu_f (\bu') =  \left(\frac{u_f}{\rho_*}\right)^{1/2}
   \  \frac{\Theta (1 + z)}{\sqrt{u'_2 + z}}.
\eeq
By equations (\ref{eq-mufsscale}) and (\ref{eq-mufp}),  we find for a fold caustic and
arbitrary surface brightness profile,
\beq
\mu_{f}^{fs} (z)
=  \left(\frac{u_f}{\rho_*}\right)^{1/2} \ \left[ {1
\over \pi} \int_{{\rm max}(-z,-1)}^{1} \rd u'_2 \ 
\frac{\Theta(1 + z)}{\sqrt{u'_2 + z}}\
\int_{-\sqrt{1-(u'_2)^2}}^{\sqrt{1-(u'_2)^2}} \rd u'_1 \
{\cS}(u'_1,u'_2)\right].
\label{eqn:mufsgen}
\eeq
For a uniform source, i.e., $\cS (\bu') = 1$, 
this simplifies to \citep{sef92, albrow1999},
\begin{equation}
\mu_{f}^{us}(z)=  \left(\frac{u_f}{\rho_*}\right)^{1/2}   \ G_0(z),
\label{eqn:muus}
\end{equation}
where 
\begin{equation}
G_n(z)\equiv \pi^{-1/2} { (n+1)! \over (n + 1/2)!}
\int_{{\rm max}(-z,-1)}^{1} {\rm d}x {(1-x^2)^{n+1/2} \over (x+z)^{1/2}} 
\Theta(1+z).
\label{eqn:gfuncn}
\end{equation}
Note that $G_0$ can be expressed as an elliptic integral.
Figure \ref{fig:fig2} shows $G_0(z)$ for $-2 \le z \le 2$.  For
small source sizes $\theta_* \ll \thetae$, the magnification $\mu_0$
of the images not associated with fold is a slowly varying
function of $\bu$ over the source, and thus
$\pi^{-1} \int_D {\rm d} \bu' S_N (\bu') \mu_0 (\bu')  = \mu_0 (\bu_{cn}).$
Therefore, the total finite source magnification
is just 
\beq
\mu_{tot}^{us}=\mu_{f}^{us} + \mu_{0,cn},
\label{eq-mtfs}
\eeq
where $\mu_{0,{cn}} \equiv \mu_0 (\bu_{cn}).$  In analogy to
the point source case (\S 2.1), we can expand $\mu_0$ about the time
of the second caustic crossing.  For a source with rectilinear motion, the 
total finite source magnification in the time domain is then,
\begin{equation}
\mu_{tot}^{us}(t)=\left( \frac{t_{f}}{\Delta t} \right)^{1/2}
G_0\left(\frac{t-\tcc}{\Delta t}\right) + \mu_{0,c} +
\dot{\mu}_{(0,c)}(t-\tcc),
\label{eqn:fsmagexp}
\end{equation}
where $t_{f}\equiv (u_{f} \te)/\sin \phi$
(effective rise time of the caustic crossing) 
and
$\Delta t \equiv  (\rho_* \te)/ \sin \phi$ is the time scale of the
caustic crossing, i.e., the time between
when the source first touches the caustic and when it straddles the caustic.
Note that $t_{f}/\Delta t = u_{f}/\rho_*$.   Figure 1(d) illustrates
the behavior of $\mu_{tot}^{us}(t)$.

\subsubsection{Finite Source Image Centroid\label{sec:fcen}}

For an extended source, the image centroid is the
position of the images weighted by both the
surface brightness  and magnification, integrated over the area of the source:
\begin{equation}
\bt_{cl}^{fs} = {
\sum_i{\int_D {\rm d} \bu S(\bu) \bt_i(\bu)\mu_i(\bu)}
\over
\sum_i{\int_D {\rm d} \bu S(\bu)}\mu_i(\bu)},
\label{eqn:tclfsgen}
\end{equation}
where the sum is over all the microimages.  The  denominator is
simply the total flux, $\bar{S} \pi \rho_*^2 \mu_{tot}^{fs}$.   
Again, for $\theta_* \ll \thetae$, the centroid $\bt_0 (\bu)$
of the images not associated with the fold varies slowly 
over the source.  This yields 
$\pi^{-1} \int_D {\rm d} \bu'  S_N (\bu') \mu_0 (\bu') \bt_0 (\bu')
 = \mu_0 (\bu_{cn}) 
     \bt_0 (\bu_{cn})$
(since
$\mu_0 (\bu)$ is also slowly varying over the
source).
Equation (\ref{eqn:tclfsgen}) then separates into two terms: 
\begin{eqnarray}
\bt_{cl}^{fs} &=& 
{ \mu_{f}^{fs}\over \mu_{tot}^{fs}}\left[ {1 \over \pi \rho_*^2\bar{S}\mu_{f}^{fs}}
\int_D {\rm d}\bu  S (\bu)\ \bt_f (\bu) \ \mu_f (\bu)\right]
\ +   \ {\mu_{0,cn} \over \mu_{tot}^{fs}} \ 
\bt_{0, {cn}} \label{eqn:tclfs1} \\
&=&
{ \mu_{f}^{fs} \over \mu_{tot}^{fs}} \  \bt_{f}^{fs}
+ {\mu_{0, {cn}} \over \mu_{tot}^{fs}}\  \bt_{0, {cn}},
\label{eqn:tclfs2}
\end{eqnarray}
where $\bt_{0,cn} \equiv \bt_0 (\bu_{cn})$.
The first term in (\ref{eqn:tclfs2}) 
is the contribution from the two images associated with
the fold caustic, while the second term is the contribution from all
unrelated images.
For convenience, we 
defined  $\bt_{f}^{fs}$ to be the factor 
within the brackets in \eq{eqn:tclfs1}.

The  term $\bt_{f}^{fs}$ will now be evaluated.
Since $\bu = \rho_* \bu' + \bu_{cn}$,
we obtain
\beq
\bt_{f}(\bu)=\bt_{f}(\bu_{cn})+ \rho_* \bt_{f}(\bu').
\eeq
Define $\bt_{f,cn} \equiv \bt_{f} (\bu_{cn}) \ \Theta (1+z),$ 
so  $\bt_{f,cn}$ vanishes when the
disc source lies completely below the fold caustic.\footnote{Technically, 
  the fold produces no images locally of $\bu_{cn}$ when $z < 0$, 
  and thus $\bt_{f} (\bu_{cn})$ 
    is not defined for $z<0$.  This apparent discrepancy can be alleviated by simply
assuming that $\bt_f(u_1,u_2)$ is a function that is defined for all $u_1,u_2$, i.e.,
$\bt_f(u_1,u_2) = (du_1 - bu_2, -bu_1)/(ad)$ for all $u_1,u_2$.}
Using (\ref{eq-mufp}),
we find 
a simple formula for the finite source image centroid,
\beq
\label{eq-theta-fs-f}
\bt_{f}^{fs}(t)
  =  \bt_{f,cn} 
   + \frac{\sqrt{u_{f} \ \rho_*}}{\mu^{fs}_{f}} \ 
\bt_{f} (\cH_{S_N}, \cK_{S_N}),
\eeq
where we have defined
\beqa
\cH_{S_N} (z)
&=& 
 \frac{1}{\pi}
\int_{{\rm max}(-z,-1)}^{1} \rd u'_2 \
\frac{\Theta(1 + z)}{\sqrt{u'_2 + z}}\
\int_{-\sqrt{1-({u'_2})^2}}^{\sqrt{1-({u'_2})^2}} \rd u'_1 \ u'_1 \
{\cS}(u'_1,u'_2),
\label{eq-cH}
\\
\cK_{S_N} (z)
&=&
 \frac{1}{\pi}
\int_{{\rm max}(-z,-1)}^{1} \rd u'_2 \ u'_2 \
\frac{\Theta(1 + z)}{\sqrt{u'_2 + z}}\
\int_{-\sqrt{1-({u'_2})^2}}^{\sqrt{1-({u'_2})^2}} \rd u'_1 \ 
{\cS}(u'_1,u'_2).
\label{eq-cK}
\eeqa

In the case of a uniform source, we obtain\footnote{In fact, $\cH_{S_N} (z) = 0$ for 
          any profile with the symmetry 
 $S_N(u'_1,u'_2)=S_N(-u'_1,u'_2)$.} $\cH_{S_N} (z) = 0$ and $\cK_{S_N}=\cK_0$, where
\beq
\cK_n (z) \equiv \pi^{-1/2} { (n+1)! \over (n + 1/2)!}
\int_{{\rm max}(-z,-1)}^{1} {\rm d}x\, x {(1-x^2)^{n+1/2} \over (x+z)^{1/2}} 
\Theta(1+z).
\label{eqn:kfuncn}
\eeq
For a uniform source, the finite source centroid of the 
two images associated with the fold is then,
\beq
\label{eq-fscentroid-us-s}
\bt_{f}^{us}
  =  \bt_{f,cn}  -  \frac{b\ \sqrt{u_{f} \ \rho_*}}{a d \ \mu^{fs}_{f}} 
 \cK_0 (z) \ \hat{{\vect{\rm \i}}}, 
\eeq
where $\hat{\vect{\rm \i}} \equiv (1,0)$.  Inserting the 
definition of $\mu^{us}_{f}$, 
this can also be written in the alternate form,
\beq
\label{eq-fsscentroid-us-alt}
\bt_{f}^{us}
  =  \bt_{f,cn}  -  \frac{b \rho_*}{a d} \frac{\cK_0 (z)}{G_0(z)}\hat{{\vect{\rm \i}}}. 
\eeq
Figure 2 shows the functions $\cK_0(z)$ and  $G_0 (z)$,
while Figure 3 depicts $\cK_0(z)/G_0(z).$  Figure 1(g) illustrates the behavior of the
image centroid (\Eq{eqn:tclfs2}) for a finite source.

\subsection{Limb Darkening\label{sec:limb}}

In this section we consider the effect of non-uniform sources on the
magnification and centroid shift near folds.  The effect of generic
surface brightnesses can be evaluated using the general integral forms
for the magnification (\Eq{eqn:mufsgen}), and the two components of
the centroid shift (Eqs. \ref{eq-cH} and \ref{eq-cK}).  We will
concentrate on a specific form for $S(\bu)$ applicable to stellar sources,
namely
\begin{equation}
S_N(\bu)=\left\{1 - \Gamma\left[1- \frac{3}{2}\left( 1-
{{|\bu - \bu_{cn}|^2}\over \rho_*^2}\right)^{1/2}\right]\right\},
\label{eqn:sbp}
\end{equation}
where $\bu_{cn}$ is the center of the source.  Here $\Gamma$ is the
limb darkening parameter, which may be wavelength dependent. This form
was originally introduced by \citet{albrow1999}, and it has the
desirable property that there is no net flux associated with the limb
darkening term.

Inserting this form for the surface brightness into \eq{eqn:mufsgen},
we recover the result of \citet{albrow1999} for the limb darkened magnification
\begin{equation}
\mu^{ld}_{f} = \mu^{us}_{f} + \Gamma \  \left(u_f
\over \rho_* \right)^{1/2} [G_{1/2}(z) - G_{0}(z)],
\label{eqn:mufld}
\end{equation}
where $G_{n}(z)$ is defined in \eq{eqn:gfuncn}, and $G_0(z)$, $G_{1/2}(z)$,
and $G_{1/2}(z)-G_0(z)$ are shown in Figure 2.

Since this form of the surface brightness profile is symmetric, 
$S_N(u'_1,u'_2)=S_N(-u'_1,u'_2)$,
the integral $\cH_{S_N}$ vanishes.  Inserting \eq{eqn:sbp} into the general form for $\cK_{S_N}$, we find
$\cK_{S_N}=\cK_0+\Gamma(\cK_{1/2}-\cK_0)$, and thus,
\beq
\label{eq-fscentroid-ld-s}
\bt_{f}^{ld}
  =  \bt_{f,cn}
   - \frac{b\ \sqrt{u_{f} \ \rho_*}}{a d \ \mu^{ld}_{f}}  [\cK_0 (z)+\Gamma(\cK_{1/2}-\cK_0)] \ \hat{{\vect{\rm \i}}},
\eeq
which can also be written,
\beq
\label{eq-fscentroid-ld-sv2}
\bt_{f}^{ld}
  =  \bt_{f,cn}
    - \frac{b \rho_*}{a d} \left[\frac{\cK_0 (z)+\Gamma(\cK_{1/2}-\cK_0)}{G_0 (z)+\Gamma(G_{1/2}-G_0)}\right] \hat{{\vect{\rm \i}}}.
\eeq
Figure 2 shows the functions $\cK_0(z)$, $\cK_{1/2}(z)$, 
and $\cK_{1/2}(z) - \cK_0 (z)$, while Figure 3 depicts the term in brackets above for 
several values of $\Gamma$. 

\subsection{Simple Linear Folds\label{sec:simple}}
 
The majority of the results presented in \S \ref{sec:analytic} (with
the exception of the assumption made in \S\S \ref{sec:fsources} and
\ref{sec:limb} that the caustic is coincident with the $u_1$-axis) are
applicable for the general case of a parabolic fold catastrophe.  In
some cases, however, it is possible to simplify the expressions
considerably and recover the more familiar
linear fold form.

To see how this limit may be reached, consider the form for the
general fold mapping in \eq{eq-le-fold}.  In this form, the source and
image plane coordinates have all been normalized by $\thetae$, and
thus are dimensionless, as are the coefficients $a,b,c,d$ (since
$\psi$ is dimensionless).  It is clear that the caustic will (in
general) be appreciably curved when one considers order unity
variations in $u_1,u_2$ (i.e.,\ on absolute angular scales of order
$\thetae$).  In other words, in the general case one can expect that
all of the coefficients ($a,b,c,d$) to be of the same order of
magnitude. Now consider variations on some smaller scale $\thetas \ll
\thetae$.  Renormalizing the angular source and image plane variables
such that $\hbu= (\thetae/\thetas) \bu$ and $\hbt=
(\thetae/\thetas)\bt$, the fold mapping can be recast in the form 
\beq
\label{eq-le-fold-2}
{\hat u}_1 = A {\hat\theta}_1 + \frac{B}{2} {\hat\theta}^2_2  +   C {\hat\theta}_1 {\hat\theta}_2,
\qquad 
{\hat u}_2 =  \frac{C}{2}{\hat\theta}^2_1 + B {\hat\theta}_1{\hat\theta}_2 + \frac{D}{2}{\hat\theta}^2_2,
\eeq
\noindent
where the relation between the coefficients are $A=a$,
$B=b(\thetas/\thetae)$, $C=c(\thetas/\thetae)$, and
$D=d(\thetas/\thetae)$.  Therefore, on scales of order $\thetas$, we
can expect $A$ to be larger than $B,C$ and $D$ by a factor
$\thetae/\thetas$.  In the Galactic microlensing case, for example,
one is typically concerned with variations on the scale of the source,
i.e.,\ scales of order $\thetas \approx \theta_*$.  In this case
$\thetae/\thetas \ga 100$.  

Let us consider the curvature of the fold caustic on scales of ${\cal O}(\thetas)$.
In general, 
for a twice continuously differentiable function
$f = f(x)$ on an open interval of ${\bf R}$,
the magnitude of the curvature at a point
$(x_0, f(x_0))$ on the graph of $f$
is given by
\beq
\label{eq-curvature}
|\hat{k}(x_0)| = \frac{f^{\prime\prime}(x_0)}{[1 + (f^\prime (x_0))^2]^{3/2}}.
\eeq
In the coordinates of
(\ref{eq-le-fold-2}), i.e.\ on scales of ${\cal O}(\thetas)$, equation
(\ref{eq-C}) yields that the fold caustic can be expressed
as a graph $(\hat{u}_1, f(\hat{u}_1)),$ where
\beq
f(\hat{u}_1) =  \hat{e} \ \hat{u}^2_1, \qquad 
\hat{e} \equiv \frac{CD-B^2}{2 A^2 D}.
\eeq
By (\ref{eq-curvature}), 
\beq
|\hat{k}(\hat{u}_1)| = \frac{2 |\hat{e}|}{[1 
\ + \ 4 \hat{e}^2 \ \hat{u}^2_1 ]^{3/2}}.
\eeq
But 
\beq
\hat{e} = \left(\frac{\thetas}{\thetae}\right) e,
\qquad \hat{u}_1 = \left(\frac{\thetae}{\thetas}\right) \ u_1.
\eeq
Hence, the magnitude of the 
curvature of the fold caustic can be expressed as
\beq
\label{eq-curvature-f}
|\hat{k}(\hat{u}_1)|  = \left(\frac{\thetas}{\thetae}\right) \frac{2 |e|}{[1
 \  + \  4 e^2 \ u^2_1 ]^{3/2}}.
\eeq
Equation (\ref{eq-curvature-f})
yields that, at the origin and on scales comparable to $\thetas$, 
the magnitude of curvature scales as $\thetas/\thetae$, and is given
by $|\hat{k}(0)|=2(\thetas/\thetae)e$.  Note that the radius of curvature at the origin 
is $|\hat{k}(0)|^{-1}$.  In Galactic microlensing $\thetas \sim \theta_*$ and $\theta_*/\thetae \ll 1$.  
Therefore it is generally the case that $|\hat{k}| \ll 1$, i.e.,\ 
the curvature of the fold caustic
is negligible on scales of order $\theta_*$, and the fold can be treated as linear.

We therefore ask what happens to the
expressions derived in the previous sections in the limit that $a
\gg (b,c,d)$.
We find that $C(u_1,u_2) \simeq 2 a^2 d u_2$, and thus the caustic,
which is defined by $C(u_1,u_2) =0$, collapses to the  
$u_1$-axis (linear fold).  The critical curve is still given by
$\theta_2 = - (b/d) \theta_1$.  There are images whenever $u_2\ge 0$,
which are located at
\beq
\bt_\pm=\left(\frac{u_1}{a},\pm \sqrt{\frac{2 u_2}{d}}\right),
\label{eq-sf-mag}
\eeq
with magnifications,
\beq
\mu_\pm = \frac{1}{2} \sqrt{\frac{u_{f}}{u_2}}.
\eeq
The centroid of the two images therefore takes on the simple form,
\beq
\bt_f
= \left(\frac{u_1}{a} , 0\right).
\eeq
 
Including the additional images, and assuming a rectilinear trajectory, 
the total magnification can be written,
\beq
\mu_{tot} (t) = \left(\frac{t_f}{t-\tcc}\right)^{1/2}\Theta (t-\tcc)  + 
 \mu_{0,c} + \dot{\mu}_{(0,c)} (t-\tcc) ,
\label{eqn:muvtime-sf}
\eeq
The centroid of all the images has the same form as before (\Eq{eqn:bxclvtime}), with
\beq
\bt_{f,c}=\left(\frac{u_{c,1}}{a},0\right)\qquad \dot{\bt}_{(f,c)}=\left(\frac{\cos\phi}{a\te},0\right).
\eeq

For finite sources, the centroid in (\ref{eq-fscentroid-us-s}) reduces to
$\bt_{f}^{fs}= \bt_{f,cn}.$
The finite source centroid then becomes
\beq
\bt^{fs}_{cl}
=
{ \mu_{f}^{fs} \over \mu_{tot}^{fs}} \  \bt_{f,cn}
+ {\mu_{0, {cn}} \over \mu_{tot}^{fs}}\  \bt_{0,cn},
\label{eq-samemag}
\eeq
This is the same result as for point sources (see \Eq{eqn:genbxcl})  if 
the point source magnification is replaced with the finite source magnification.  

Equations (\ref{eq-sf-mag})-(\ref{eq-samemag}) should hold whenever (1) $a\gg(b,c,d)$, (2)
one is not too close to a higher-order catastrophe, i.e.,\ a cusp, and (3) there are no other nearby 
folds.  

\subsection{Some Applications}

We can use the results from the previous sections to derive some
generic results about the astrometric behavior near folds.  We first
consider the magnitude of the astrometric jump when the source crosses
a caustic.  It is clear that for a point source, the maximum centroid
shift is $\dtj\equiv|\bt_{f,c} - \bt_{0,c}|$, where $\bt_{f,c}$ is the
point on the critical curve where the images merge.  Although $\dtj$
ultimately depends on the source trajectory and the topology of the
lens, we can typically expect that $\dtj$ is ${\cal O}(\thetae)$.  For
the finite source case, the difference between the centroid position
just before ($z \sim -1$) and just after ($z \sim 1$) the caustic
crossing is $\dtj^{fs} \simeq (\mu_{f}^{fs}/\mu_{tot}^{fs})\dtj$.  For
a uniform source, this is
\begin{equation}
\dtj^{us} \simeq \left[ 1 + \sqrt{ \rho_* \over u_{f}} \frac{\mu_{0,cn}}{G_0}\right]^{-1} \dtj.
\label{eqn:fsjump}
\end{equation}
Adopting typical parameters, $\mu_{0,cn} \sim 1$, $u_{f}\sim 1$,
and $G_0\sim 1$, we find the fractional change from the point source
case to be $(\dtj^{fs}-\dtj)/\dtj \sim
-3\%(\rho_*/10^{-3})^{1/2}$.

We now consider the magnitude of limb darkening effect on the
magnification and centroid shift relative to the uniform source case.
From equations (\ref{eq-mtfs}) and (\ref{eqn:mufld}), the fractional
difference between the limb-darkened and uniform-source magnification
is,
\begin{equation}
\delta \mu_{ld} \equiv 
{{\mu_{tot}^{ld}- \mu_{tot}^{us}} \over {\mu_{tot}^{us}}}=
\Gamma {{G_{1/2}-G_0}\over G_0}\left[ 1 + \sqrt{ \rho_* \over u_{f}} \frac{\mu_{0,cn}}{G_0}\right]^{-1}.
\label{eqn:deltamu}
\end{equation}
Early in the caustic crossing ($z < 1$), the factor
$\sqrt{\rho_*/u_{f}}(\mu_{0,cn}/G_0)$ is small compared to unity, and thus
$\delta \mu_{ld} \sim \Gamma (G_{1/2}-G_{0})/G_{0}$.  The magnitude of
$(G_{1/2}-G_{0})/G_{0}$ is $\le 20\%$ for the majority ($z\la 0.8$) of
the caustic crossing.  Near the end of the caustic crossing ($z \sim
1$) the term in brackets begins to dominate as $G_0 \rightarrow 0$,
and thus $\delta \mu_{ld} \sim \Gamma
\mu_{0,cn}^{-1}\sqrt{u_{f}/\rho_*}(G_{1/2}-G_{0})$, which goes to zero as
$z \rightarrow 1$. Thus for typical values of $u_f, \rho_*$, and
$\mu_{0,cn}$, the fractional difference from a uniform source is $\la
0.2\Gamma$ for the majority of the caustic crossing.  See Figure 3.

The difference in the centroid due to limb darkening is
\beq
\Delta\bt_{cl}^{ld}
=
\bt_{cl}^{ld} - \bt_{cl}^{us}
= \delta \mu_{ld}  \  \frac{\mu_{0,cn}}{\mu_{tot}^{ld}}\ (\bt_f - \bt_0).
\label{eqn:deltatld}
\eeq
Figure 3(c) shows the prefactor $\delta \mu_{ld} {\mu_{0,cn}}/{\mu_{tot}^{ld}}$ for 
$\mu_{0,cn}=4$, $u_f=1$, and several values of $\rho_*$. 
To assess the detectability of the deviation of the centroid shift
from the uniform source case due to limb darkening in Galactic bulge
microlensing events, we now make a crude estimate for the maximum
magnitude of $\Delta\bt_{cl}^{ld}$ by adopting typical parameters.  At
$t=\tcc$, we have that $|\bt_f - \bt_0|= \dtj$, and is maximized.  We
will assume that $\dtj \sim \thetae \sim 300 \muas$.  We have just
argued that $\delta \mu_{ld}\la 0.2 \Gamma$.  Near the end of the
caustic crossing (where $\delta \mu_{ld}$ is maximized),
$\mu_{tot}^{ld} \sim \mu_0$.  Inserting these values into
\eq{eqn:deltatld}, we find
\begin{equation}
|\Delta \bt_{cl}^{ld}|_{max} \sim (60\Gamma) \,\muas .\qquad ({\rm Bulge\,
 Lenses}).
\label{eqn:ldmax}
\end{equation}
For typical values of $\Gamma \sim 0.5$ in the optical, $|\Delta
\bt_{cl}^{ld}|_{max} \la 30 \muas$, which is only a factor of $\sim 3-6$
larger than the sensitivity expected from {\it SIM}.  Note that this is the {\it maximum}
centroid shift; inspection of Figure 3(c) reveals that $|\Delta
\bt_{cl}^{ld}|$ is considerably smaller than this maximum for the
majority of the caustic crossing. Therefore, detection of the
astrometric effects of limb-darkening will likely be challenging, at
least for Galactic bulge lensing events, with the currently planned
accuracy for SIM. On the other hand, the fact that the astrometric effects
of limb darkening are small implies that limb-darkening can
generally be ignored.  Thus predictions for the signatures of other
effects, such as the finite source effect itself, are robust.  In
other words, measurements generally should not be compromised
by an unknown amount of limb-darkening.

Consider two observers that are not spatially coincident.  The source
trajectories as seen by the two observers will be displaced relative to each
other by an amount
$\delta \bu$, the magnitude of which is $|\delta \bu| = a_{\oplus s}/
\tilde r_{\rm E}$, where $a_{\oplus s}$ is the component of the separation
between the two observers perpendicular to the line-of-sight, 
and $\tilde r_{\rm E}=\thetae D$ is the
Einstein ring radius projected to the observer plane.  If $\delta \bu$
is sufficiently large, then both the photometric and astrometric
behavior of the event will be measurably different between the two observers, an
effect commonly known as parallax.  Since $a_{\oplus s}$ is known,
this observed difference can be used to infer $\tilde r_{\rm E}$,
providing additional physical constraints on the properties of the
lens.  For example, when combined with a measurement of $\thetae$ from
the astrometric centroid shift itself, the mass of the lens can be
determined, $M=(c^2/4G)\tilde r_{\rm E} \thetae$.  For a given $|\delta \bu|$, parallax
effects are typically largest near caustics, because the 
magnification and the centroid vary rapidly with respect to source position.   
Thus caustic
crossings are ideal for use in measuring parallax effects \citep{hw1995,ga1999}.   
If $|\delta \bu|$ is sufficiently small that 
the behavior near the fold caustic crossing for both observers can be
described by the same local expansion of the lens mapping, than the results derived here 
can be used to fit the astrometric and photometric behavior for both observers and
derive the parallax effects without regard to the global behavior of the lens.
We note that, for generic fold caustics, any significant
displacement $\delta \bu$, regardless of the its orientation relative to the caustic, 
will result in a difference in the observable behavior.  However,
for linear fold caustics (\S \ref{sec:simple}), only displacements perpendicular
to the caustic will result in significant differences in the magnification and centroid.  Thus, 
for generic linear fold caustics, only a projection of $\tilde r_{\rm E}$ is measurable
from the local caustic behavior \citep{gg2002}.

During a fold caustic crossing the source is
resolved, altering both the photometric and astrometric behavior with
respect to a point source (\S \ref{sec:fsources}).  The photometric
behavior near a caustic crossing depends on $\rho_*$, the source size
in units of $\thetae$, whereas the astrometric behavior depends on
actual angular size of the source $\theta_*=\rho_*\thetae$.\footnote{Recall 
that, in order to covert to astrometric observables, all angular variables must
be multiplied by $\thetae$.}
Thus,
while fitting to the photometric data near a caustic crossing does not
yield the angular size of the source, fitting to the astrometric data
near a caustic crossing does.  Therefore it is possible, in principle, to
measure the angular radius of the source star of a caustic-crossing
binary-lens event by fitting a few astrometric measurements taken during
the caustic crossing to the expressions we have derived 
for the local astrometric behavior (see \S \ref{sec:fcen} and Eq.\ \ref{eqn:tclfs2}).
In practice, however, this is complicated
by the fact that, for linear fold caustics, only the degenerate
combination $\theta_*/\sin\phi$ can be measured, where $\phi$ is the
angle of the trajectory with respect to the caustic.  For general fold
caustics, the degeneracy between $\theta_*$ and $\sin\phi$ is broken
(see \S \ref{sec:fcen} and Eq.\ \ref{eq-fsscentroid-us-alt}), however, for many cases the simple linear
fold will be applicable. Thus, is in order to determine
$\theta_*$ separately, the global geometry must be generally
specified, which can be accomplished using the global photometric light curve.

\section{A Worked Example: Binary Lensing Event OGLE-1999-BUL-23}

In this section we numerically calculate the expected astrometric
behavior for a binary-lens fold caustic crossing, and compare this
with the analytic results from the previous sections.  We do this in
order to verify our expressions and also to explore the accuracy with
which our (necessarily) approximate results reproduce the exact
behavior.  For definiteness, we will calculate the expected
astrometric behavior for the photometrically well-observed
caustic-crossing binary lens event {\event} \citep{albrow2001}.  This
has the advantage that, up to an orientation on the sky and subject to
small errors in the inferred parameters, the astrometric behavior can
be essentially completely determined from the photometric solution,
including the size of $\thetae$ and the effects of limb-darkening.

\subsection{Formalism and Procedures}

For a system of $N_l$ point masses located at positions $\bt_{l,j}$,
and no external shear, the lens equation (\Eq{eqn:lenseq}) takes the form
\begin{equation}
\bu =\bt - \sum_{j=1}^{N_l} m_j
{{\bt - \bt_{l,j}} \over {|\bt - \bt_{l,j}|^2}},
\label{eqn:blenseq}
\end{equation}
where $m_j$ is the mass of the $j$-th lens in units of the total mass.  
Note that angles in \eq{eqn:blenseq} are normalized
to the $\thetae$ for the total mass of the system.
For $N_l=2$, the
lens equation is equivalent to a fifth-order polynomial in $\bt$, thus
yielding a maximum of five images.  All of the image positions for a
given point on the source plane can be found numerically using any
standard root finding algorithm.  Then the individual magnifications,
total magnification and centroid of these images can be found using
equations (\ref{eqn:magoneimage}), (\ref{eqn:mutot}), and
(\ref{eqn:cl}).

For a finite source size, it is necessary to integrate over the area
of the source (\Eq{eqn:fsgen}).  This can be difficult to do
numerically in the source place near the caustics, due to the
divergent magnification.  A more robust method is inverse ray
shooting.  This works as follows.  The image plane is sampled
uniformly and densely, and at each $\bt$, \eq{eqn:blenseq} is used to
find the corresponding $\bu(\bt)$.  The local ratio of the density of
rays in $\bt$ to the density of rays in $\bu$ is the local
magnification.  Thus one can create a map of $\mu(\bu)$, the
magnification as a function of $\bu$.  Similarly, one can determine
the astrometric deviation by sampling in $\bt$, using \eq{eqn:blenseq}
to determine $\bu(\bt)$, and then summing at each $\bu$ the values of
$\bt(\bu)$.  The astrometric deviation at $\bu$ is then the summed
values of $\bt(\bu)$, weighted by the local magnification.  Thus one
creates two astrometric maps, for each direction.  In practice,
inverse ray shooting requires one to bin the rays in the source plane,
with the resolution of the maps being determined by the size of the
bin, and the accuracy determined by surface density of rays in the
image plane relative to the (unlensed) surface density of rays in the
source plane.  The advantage of inverse ray shooting is that the
procedure conserves flux, therefore the maps can be convolved with any
source profile to produce the finite source photometric and
astrometric behavior for arbitrary source size and surface brightness
profile.

To apply this method to predict the detailed photometric and
astrometric behavior of the caustic crossings for \event, it is
essential that the resolution of the astrometric and photometric maps
be considerably smaller than the source size $\rho_*$.  We will be
using resolutions of $10^{-4}\thetae$, which corresponds to $0.034
\rho_*$ for \event.  This is sufficient to accurately resolve the
source.  We sample the image plane with a density of $5\times 10^9
\thetae^{-2}$, corresponding to an Poisson error per resolution
element of $\sim 14\% \mu^{-1}$.  Since there are $\sim 2700$
resolution elements per source size, the total Poisson error is always
$< 1\%$, considerably smaller than any of the effects we will be
considering.

\subsection{Global Astrometric Behavior}

Before studying the detailed behavior near the photometrically
well-covered (second) caustic crossing of \event, we first analyze the
global astrometric behavior of the entire event.  We specify the
binary-lens topology and source trajectory using the parameters of the
best-fit solution with limb darkening (see Table 2 of
\citealt{albrow2001}).  In Figure 4 we show various aspects of the
inferred lensing system and the event itself.  The best-fit lens
system has a wide topology, with two well-separated caustic curves, one near
the position of each lens.  The mass ratio of the two lenses is
$q\equiv m_1/m_2 = 0.39$, and they are separated by $2.42\thetae$.
The solution has the source crossing the caustic associated with the
least massive lens, which we will call the secondary caustic.

Figure 4(a) shows the photometric light curve centered on the event.
There are two fold caustic crossings, separated by $\sim 10$ days; the
second crossing was densely covered photometrically by
\citet{albrow2001}, allowing them to determine not only the source size
$\rho_*$, but also limb-darkening coefficients $\Gamma$ in each of two
different photometric bands: $\Gamma_I=0.534$ ($I$-band) and
$\Gamma_V=0.711$ ($V$-band).  This allows us to predict the
astrometric behavior including finite source and limb-darkening
effects.  Furthermore, by combining a measurement of $\rho_* \equiv
\theta_*/\thetae$, with a determination of the angular size of the
source $\theta_*$ from its color and magnitude, \citet{albrow2001}
measured the angular Einstein ring radius of the lens to be $\thetae =
(634\pm 43) \muas$. Therefore we can determine the absolute scale of the
astrometric features and assess their detectability by comparing them
with the expected accuracy of upcoming interferometers.

Figure 4(b) shows the caustics and critical curves, as well as the
trajectory of the images of the source.  One image is always near the
most massive lens; this image has little effect on the resulting
astrometric deviation other than a small net offset along the binary
axis.  Figure 4(c) shows a close-up of the region near the secondary
caustic, along with the image centroid $\btcl$.  The components
parallel and perpendicular to the binary axis are shown in Figures
4(e,d).  Finally, Figure 4(f) shows $\delta\btcl$, the centroid
relative to the unlensed source position $\bu$.  In Figures 4(c-f),
the large, ${\cal O}(\thetae)$ jumps that occur when the source
crosses the caustic are evident.

\subsection{The Second Fold Crossing}

We now focus on the photometric and astrometric behavior near the
second caustic crossing for \event.  In Figure 5(a) we show the
caustic geometry and source trajectory near this crossing.  The source
passed $\sim 0.02~{\rm mas}$ from a cusp.  On the scale of the source
size, $\theta_*=1.86\muas$, the curvature of the caustic can be
neglected.

In Figure 5 we show the behavior near the second caustic crossing.
Figure 5(c) shows the photometric behavior for two days centered on
the caustic crossing, for the assumption of a uniform source, and a
limb-darkened source with surface brightness profile given in
\eq{eqn:sbp}, with $\Gamma_I=0.534$ and $\Gamma_V=0.711$.  Note the
similarity of the shapes of the uniform source and limb-darkened
light curves with the analytic forms $G_0$ and $G_{1/2}$ presented in
Figure 2(a).  We show our prediction for the  astrometric behavior as a function
of time for the components parallel and perpendicular to the binary
axis in Figures 5(d) and (e), respectively
[compare with Figure 1(e,f)].  We show both the
instantaneous discontinuous jumps for a point-source centroid, 
along with the continuous centroid curves for finite sources.
The limb-darkened and uniform source astrometric curves are extremely
similar.  In Figure 5(f) we show the predicted total astrometric 
behavior for the
same time span.  Notice how well the form of Figure 5(f) compares
with Figure 1(g), which was obtained from our analytic forms.
Figure 6 shows the same curve as Figure 5(f), except
we have rotated the axes by $\sim 55^\circ$, shifted the origin to the
image of the caustic crossing, and stretched the axes for visibility.
In Figure 6, notice how the predicted shaped for the finite source centroids
are smoothly rounded off near the bottom and has a sharp turn
near the top.

Although the difference between the finite source and point source
curves is quite substantial, the difference between the uniform source
and limb darkened sources is very small.  This is more clearly
illustrated in Figure 7(a,b), where we show the two components of this
difference,
$\Delta\bt_{cl}^{ld}=\bt_{cl}^{ld}-\bt_{cl}^{us}$, for about
four source radius crossing times centered on the crossing, i.e.\
$|z|<2$.  The form of $\Delta\bt_{cl}^{ld}$ is
very similar to the analytic expectation [see \Eq{eqn:deltatld} and
Figure 3(c)].  Furthermore, the two components of
$\Delta\bt_{cl}^{ld}$ are essentially perfectly (anti-) correlated,
implying that the difference is essentially one-dimensional.  
This can be seen best in Figure
7(c), where we plot $\Delta\theta_{cl,1}^{ld}$ versus
$\Delta\theta_{cl,2}^{ld}$.  Panel (d) is the same as panel (c), except we
have rotated and stretched the axes.  The maximum deviation is $\sim
(60 \Gamma)\muas$, in satisfying agreement with the rough expectation
(\Eq{eqn:ldmax}).

\section{Summary and Conclusion}

We have presented a detailed study of gravitational lensing near fold
catastrophes, concentrating on the regime where the individual images
are unresolved, i.e.\ microlensing.  By Taylor expanding the scalar
potential $\psi$ in the neighborhood of a fold up to third order in
the image position, one can obtain a generic form for the lensing map
near a fold.  Beginning with this mapping, we derive the local
lensing properties of a source in the vicinity of the fold caustic.
Approximating the critical curve by its tangent line at the origin, we
find that the caustic is locally a parabola.  On one side of the
parabola, the fold lenses a nearby source into two images; on the
other side of the parabola, there are no images.  We derive the image
positions and magnifications as a function of the position of the
source.  We find that the magnifications of the two images are equal,
and recover the well-known result that the magnification is inversely
proportional to the square root of the distance to the caustic.  We
show how this holds for parabolic caustics (as well as linear
caustics), provided that the `vertical' distance from the caustic is
used.

Assuming a rectilinear source trajectory, and allowing for the
existence of slowly- and smoothly-varying images not associated with
the fold caustic, we derive analytic expressions for the total
magnification and image centroid (center-of-light) as a function of
time.  

We then consider how the photometric and astrometric behavior is
altered in the presence of a finite source size.  We derive
semi-analytic expressions for the magnification and centroid as a
function of time for both a uniform source, and limb-darkened source.
Along the way we derived expressions that can be used to evaluate the
photometric and astrometric behavior near a fold for a source
with arbitrary surface brightness profile. 

We then show how and under what conditions the generic parabolic fold
reduces to the more familiar linear fold.  We derive simplified
expressions for the individual and total image positions and
magnifications near a linear fold.

We used some of our analytic results to derive a few generic properties
of microlensing near folds.  In particular, we derive and evaluate
expressions for the magnitude of the centroid jump that occurs when a
finite source crosses a fold relative to the point source jump, and
the magnitude of the effect of limb darkening on both the photometric
and astrometric behavior.  Notably, we predict, for Galactic bulge
lensing events, the shape of the centroid due to finite sources
with uniform and limb darkening surface brightness profiles.
We also find for Galactic bulge lensing
that the effect of limb darkening on the image centroid
near a fold is quite similar to the uniform source case, making
the limb darkening effect difficult to detect by
the currently planned accuracy for the instrumentation of {\it SIM}.
We discussed how our formulae can be used to fit both photometric and astrometric 
data sets near fold caustic crossings 
and thus used to derive such properties as the angular size of the source and the 
microlensing parallax.

Finally, we numerically calculate expected astrometric behavior of
the photometrically well-observed Galactic bulge binary lensing event
{\event} \citep{albrow2001}, finding excellent agreement with our
analytic predictions.

Caustics are ubiquitous in gravitational lenses, and the most common
type of caustic is the fold.  Caustics play an especially important
role in microlensing, as the rapid time variability of the total image
magnification allows the possibility of detailed studies of
the source and lens.  In the future, we can expect that time-series
photometric measurements will be supplemented by time-series
{\it astrometric} measurements of the center-of-light of microlens systems.
This paper presents the most thorough and comprehensive study of the
photometric and astrometric behavior of gravitational microlensing near
fold caustics to date. The results should prove useful to those studying
microlens systems with caustics:  The analytic expressions derived here
can be used to fit fold caustic crossings observed both
photometrically and astrometrically, gain some insight into more
complicated numerical studies, and establish predictions
for the feasibility of future observations.

\acknowledgements

We would like to thank the referee, Eric Agol, for several helpful suggestions
that led to a much improved manuscript.
B.S.G. was supported in part by NASA through a Hubble Fellowship grant
from the Space Telescope Science Institute, which is operated by the
Association of Universities for Research in Astronomy, Inc., under
NASA contract NAS5-26555.  A.O.P.  was supported in part by an Alfred
P. Sloan Research fellowship and NSF Career grant DMS-98-96274.

\clearpage

\begin{figure}
\plotone{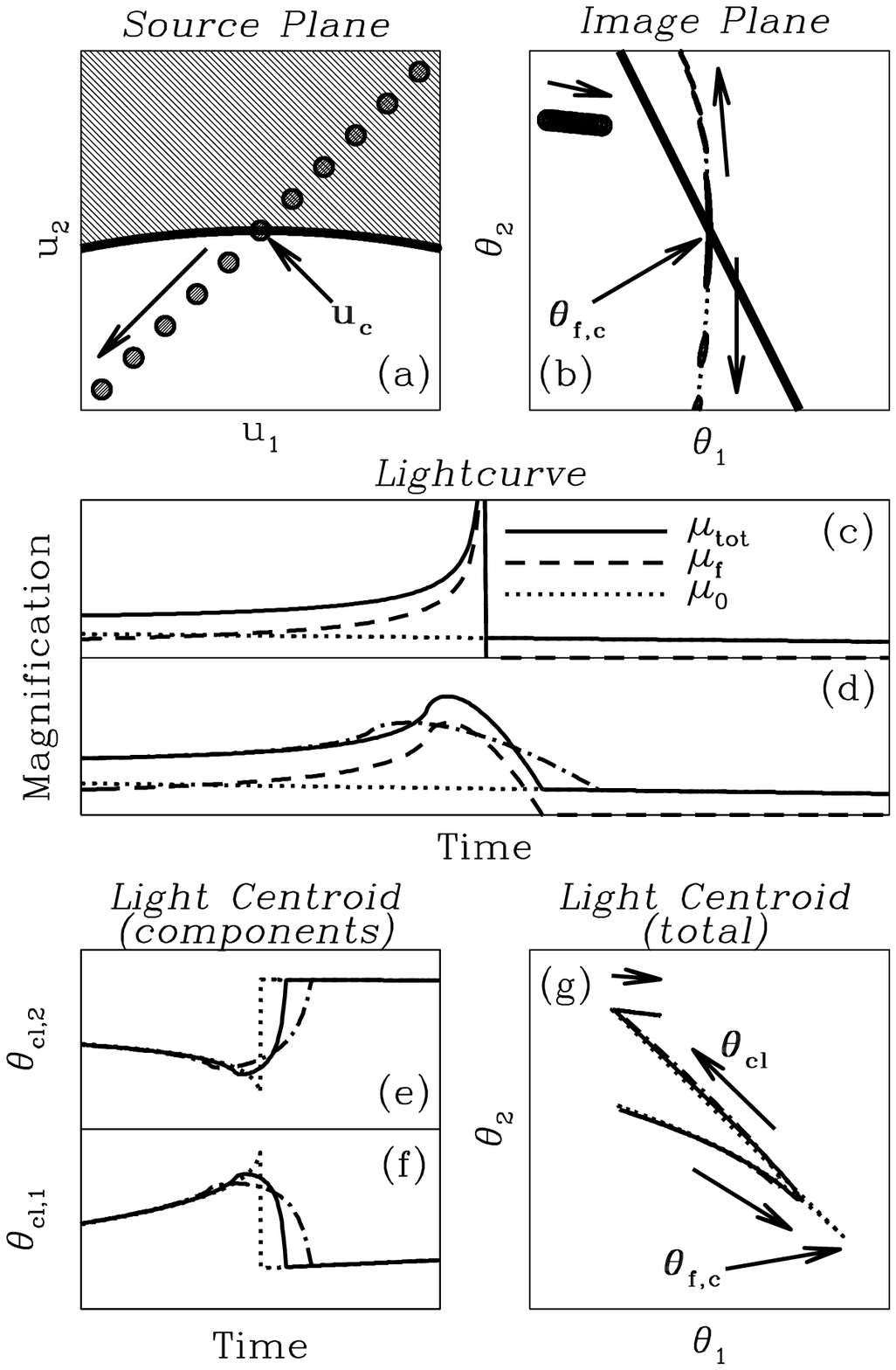}
\vskip-0.5in
\caption{ 
An illustration of the basic properties of astrometric and
photometric microlensing near folds.  (a) Filled circles represent the
source at various times.  The solid line is the fold.  The source
crosses the caustic at $\bu_c$.  (b) The elongated shapes are the
images corresponding to the source at the positions in panel (a).  The
point $\bt_{f,c}$ is the image of $\bu_c$, and is where the two extra
images appear.  The third image on the nearly horizontal trajectory
represents the centroid of all images unassociated with the fold. (c)
The magnification as a function of time for a point source.  The solid
line is the total magnification $\mutot$, the dotted line is the
magnification of all images unassociated with the caustic $\mu_0$, and
the dashed line is the magnification $\mu_{f}$ of the two images
created in the fold crossing.  (d) Same as (c), except for a finite
uniform source.  The dashed-dot line shows
the total magnification for a source size that is two times larger.  
(e) The $\theta_{cl,1}$-component of the
centroid shift as a function of time.  Dotted line is for a
point-source, solid line for a finite uniform source size.  Dashed-dot
line is for a source size that is two times larger. (f) The
$\theta_{cl,2}$-component of the centroid shift.  (g) The path of the
centroid of light of all the images, $\btcl$.  Line types are as in panels (e,f).  }
\label{fig:fig1}
\end{figure}

\begin{figure}
\plotone{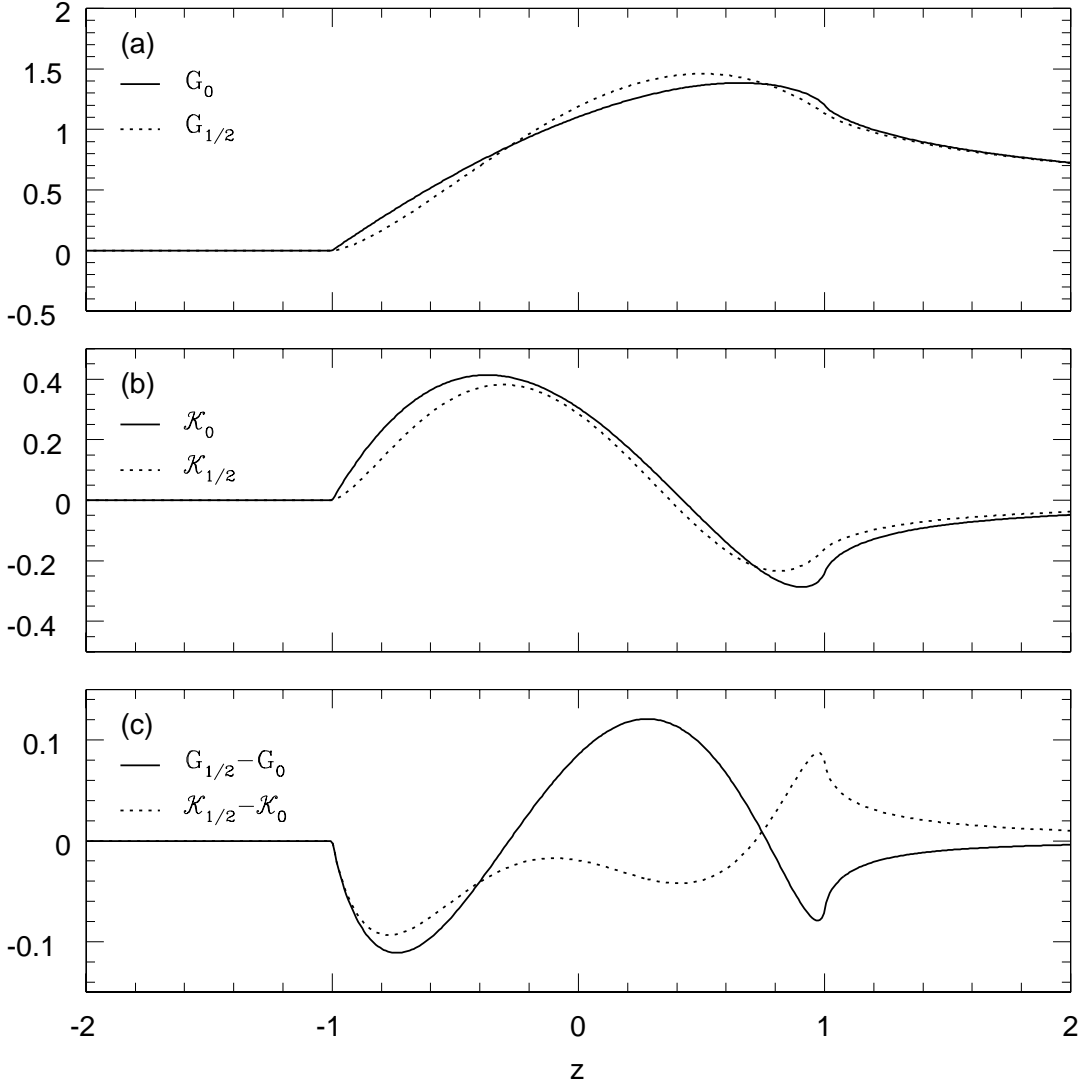}
\caption{ Basic functions which describe the photometric and
astrometric behavior of finite sources near a fold, as a function of
the distance $z$ from the fold in units of the dimensionless source
size $\rho_*$. (a) The basic functions for the photometric behavior.
The solid line shows $G_0$, whereas the dotted line shows $G_{1/2}$. 
 (b) The basic functions for the astrometric
behavior.  The solid line shows $\cK_0$, whereas the dotted
line shows $\cK_{1/2}$.
(c) The photometric ($G_{1/2}-G_{0}$) and 
astrometric ($\cK_{1/2}-\cK_{0}$) limb darkening functions.  See text.}
\label{fig:fig2}
\end{figure}

\begin{figure}
\plotone{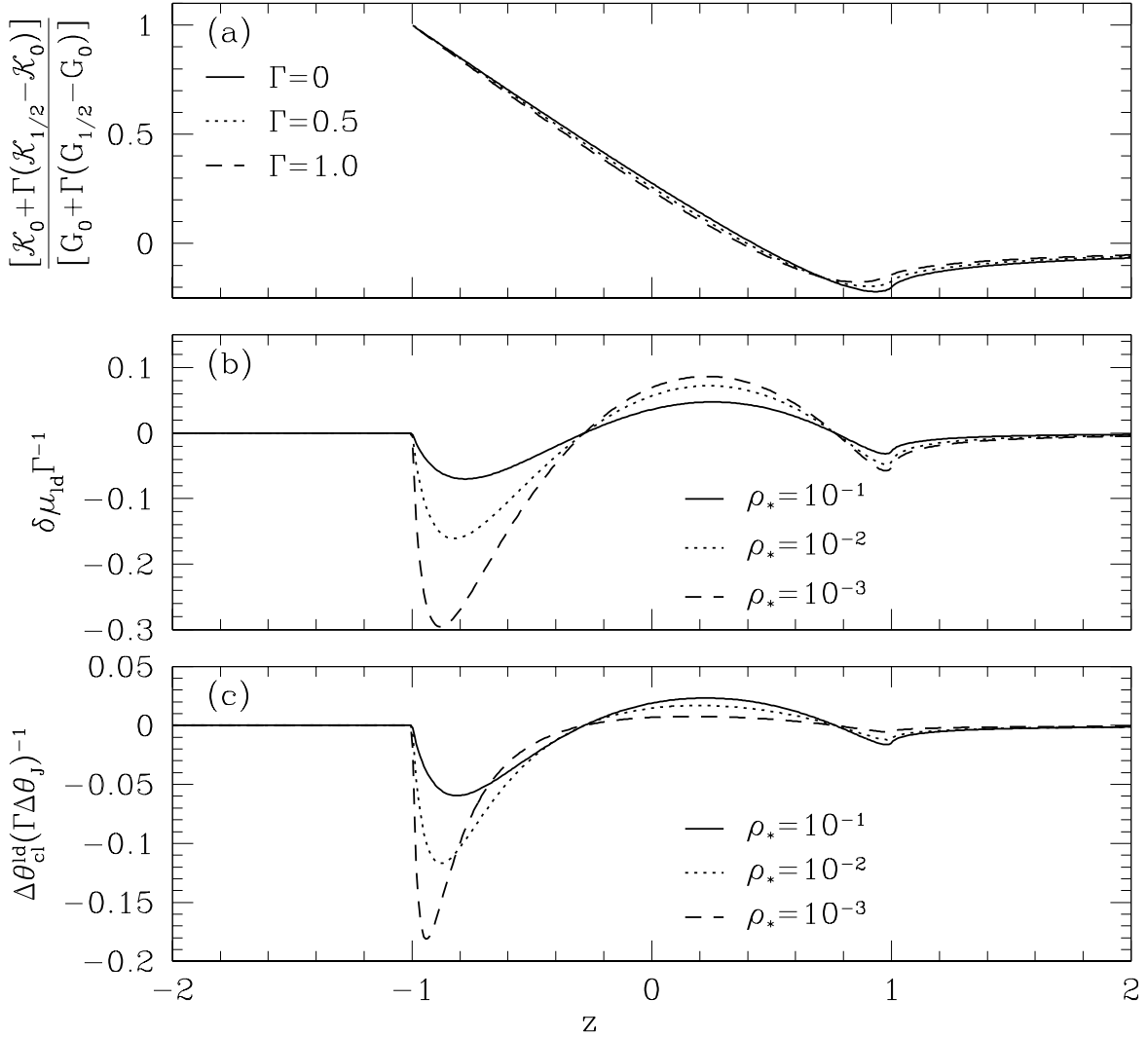}
\caption{(a) The function $[\cK_0 +\Gamma(\cK_{1/2}-\cK_{0})]/[G_0
+\Gamma(G_{1/2}-G_{0})]$ as a function of $z$, for $\Gamma=0, 0.5$,
and $1.0$, where $G_0$, $G_{1/2}$, $\cK_0$, and $\cK_{1/2}$ are shown
in Figure 2, and $\Gamma$ is a limb-darkening parameter.  (b) The fractional
difference $\delta\mu_{ld}$ between the limb-darkened and uniform
source magnifications, normalized by $\Gamma$, for several source
sizes $\rho_*$.  We have assumed a caustic scale $u_f=1$, and
magnification outside the caustic of $\mu_{0,cn}=4$.  (c) The absolute
magnitude of the difference in the centroid shift due to limb
darkening $\Delta\theta_{cl}^{ld}$, normalized by $\Gamma \dtj$, where
$\dtj$ is magnitude of the point-source astrometric `jump' when the
source crosses the caustic.  See text.  }
\label{fig:fig3}
\end{figure}

\begin{figure}
\plotone{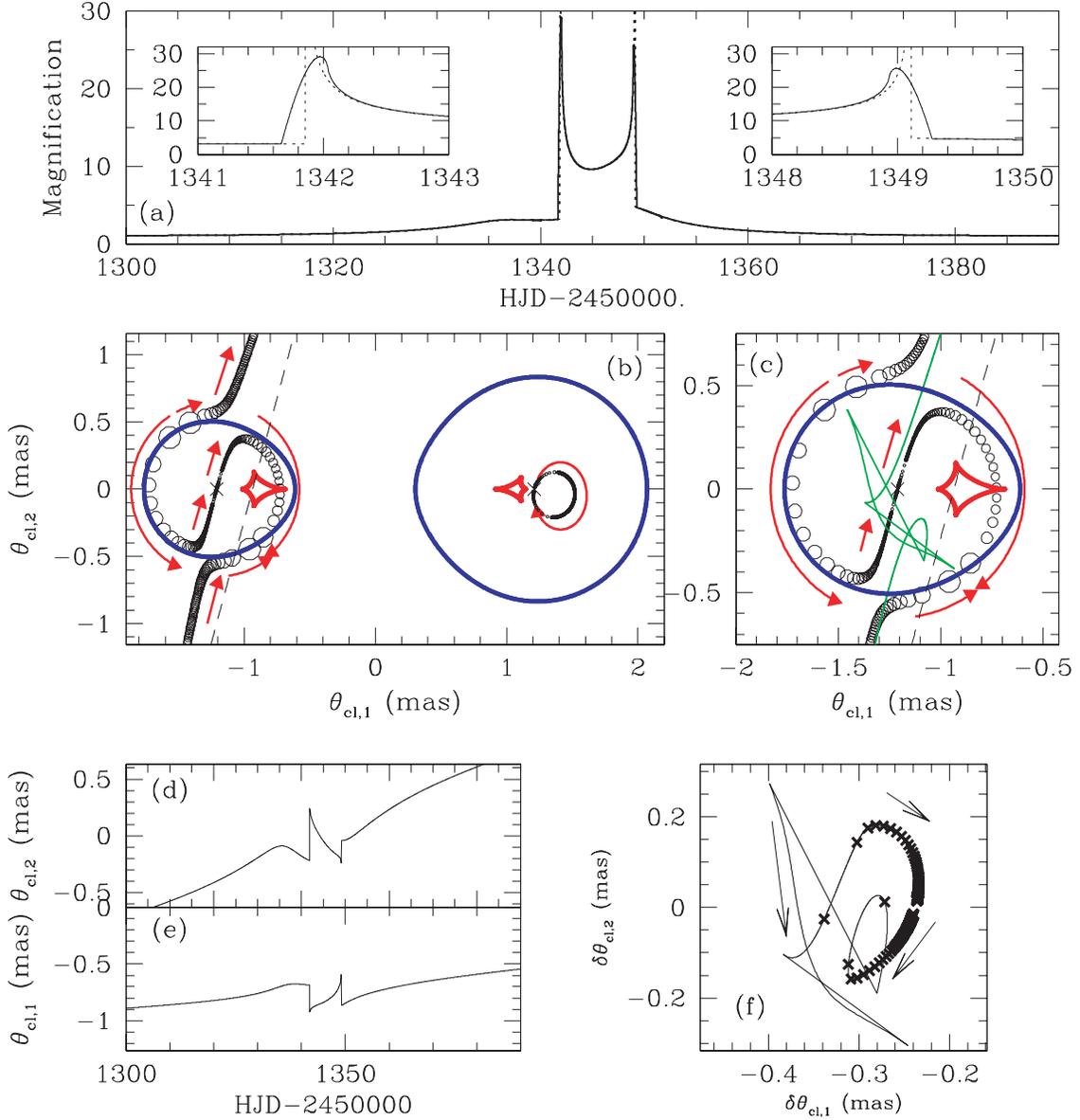}
\caption{ Global photometric and astrometric behavior of the
binary-lens event OGLE-1999-BUL-23.  (a) The light curve (magnification
as a function of time) for the best-fit model.  Solid line is for the
finite source, whereas the dotted line is for a point source. The
insets show detail near the two caustic crossings.  (b) The critical
curves (blue ovals), caustics (red cuspy curves), images
(circles), and source trajectory (dashed line).  The X's denote the
position of the two masses, the arrows give the directions of motion
of the images, and the size of the circles are proportional to the
magnification of the image.  (c) Detail near the caustic crossing.
The green lines shows the position of the centroid
$\vect{\theta}_{cl}$ of the five images relative to the lens.  (d,e)
the two components of $\vect{\theta}_{cl}$ as a function of time.  (f)
The solid line shows the centroid relative to the source position
$\delta\vect{\theta}_{cl}$.  The X's show $\delta\vect{\theta}_{cl}$
at fixed intervals of 20 days.  }
\label{fig:fig4}
\end{figure}

\begin{figure}
\plotone{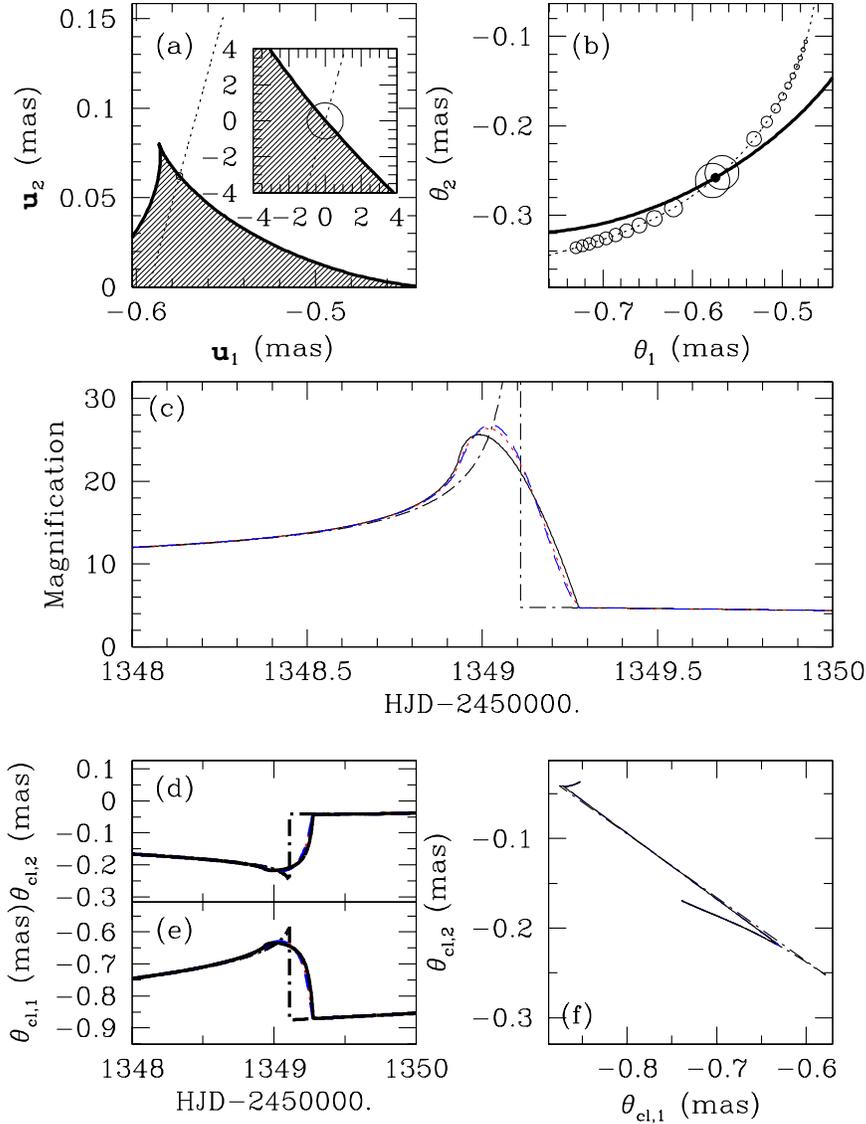}
\caption{ The photometric and astrometric behavior near the second
caustic crossing of OGLE-1999-BUL-23.  (a) The heavy solid line shows
the caustic, while the dashed line shows the source trajectory.  The
interior of the caustic is shaded.  The inset shows the detail near
the caustic crossing, in units of the source size, shown as a
circle. (b) The heavy solid line is the critical curve, and the
circles show the positions of the two images associated with the
caustic crossing at fixed intervals of 4.8 hours.  The size of the
circles is proportional to the logarithm of the magnification.  (c)
The magnification near the second caustic crossing as a function of
time.  The solid lines is for a uniform source, dotted line for a
limb-darkened source in the $I$-band, dashed line for the $V$-band,
and dashed-dot line is for a point source.  (d,e) The two components
of the centroid $\vect{\theta}_{cl}$ in mas as a function
of time.  Line types are the same as (c).  (f) The centroid
$\vect{\theta}_{cl}$. }
\label{fig:fig5}
\end{figure}

\begin{figure}
\plotone{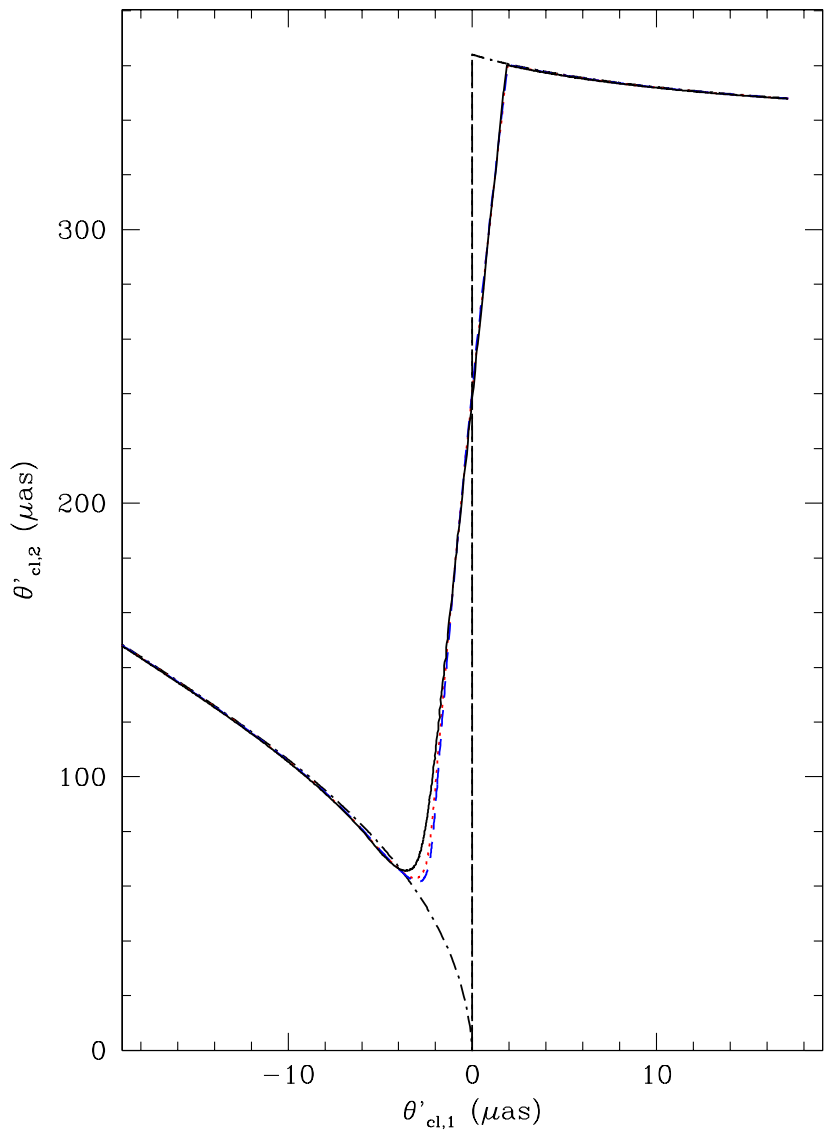}
\caption{ Detail of the centroid shift $\vect{\theta}_{cl}$ near the
second caustic crossing of OGLE-1999-BUL-23 in $\muas$.  This is
the same as panel (f) in Figure 5, except the axes have been rotated
by $\sim 55^\circ$, and the origin has been translated to the image
position of the caustic crossing point.  Note the extreme asymmetry in
the scales of the two axes.  The solid line is 
for a uniform source, dotted line for a
limb-darkened source in the $I$-band, dashed line for the $V$-band,
and dashed-dot line is for a point source.  }
\label{fig:fig6}
\end{figure}

\begin{figure}
\plotone{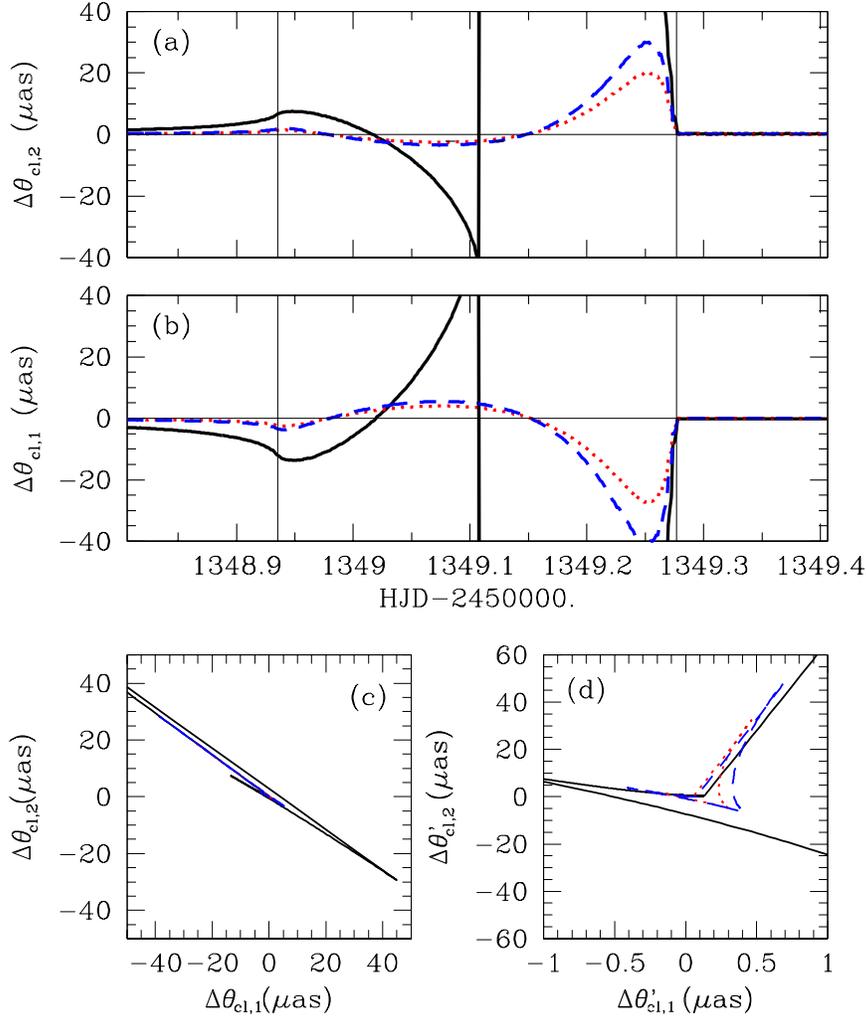}
\caption{
(a,b) The two components of the astrometric offset due to limb darkening 
$\Delta\bt_{cl}^{ld}$ relative
to a uniform source as a function
of time.  The dotted red line is for the $I$-band, where as the dashed line is for the
$V$-band.  The two components are parallel (a) and perpendicular (b) to the binary
axis. The vertical lines show (from left to right) 
the time when the source first touches, straddles, and
last touches the caustic.  The solid lines is the 
offset from a point source.  (c) The astrometric
offset $\Delta\bt_{cl}^{ld}$.  Line types are as in panels (a) and (b).  
The deviations from the uniform
source are essentially one-dimensional.  (d) The same as panel (c), 
except that the axes have
been scaled and rotated by $\sim 55^\circ$.  
Note the extreme asymmetry in the scales of the
axes.  The maximum absolute deviation 
due to limb darkening is small, $\la 50 \mu{\rm as}$.
}
\label{fig:fig7}
\end{figure}

\end{document}